\documentclass[fleqn,usenatbib]{mnras}
\usepackage[T1]{fontenc}
\usepackage{ae,aecompl}
\usepackage{graphicx}	
\usepackage{amsmath}	
\usepackage{amssymb}	
\usepackage{tikz}
\usepackage{times}
\usepackage[normalem]{ulem}
\usepackage{multirow}
\usepackage{tabularx}
\usepackage{bm}            
\usepackage{url}
\usepackage{color,colortbl}
\usepackage{hyperref}

\pdfoutput=1

\definecolor{Gray}{gray}{0.9}

\newcommand{\kms}{\ensuremath{~\mathrm{km~s^{-1}}}}

\newcommand{\Msun}{\ensuremath{~\mathrm{M}_\odot}}

\newcommand{\tinf}{\ensuremath{t_\mathrm{infall}}}
\newcommand{\tquench}{\ensuremath{t_\mathrm{q;\ 90}}}
\newcommand{\Dtquench}{\ensuremath{\Delta t_\mathrm{q;\ 90}}}

\newcommand{\reffig}[1]{Fig. \ref{#1}}

\newcommand{\eagle}{EAGLE}
\newcommand{\auriga}{Auriga}

\usepackage{color}
\definecolor{mycolor}{rgb}{.0,.3,1.}

\begin{document}



\title[Satellite Infall Times using NNs]{Determining Satellite Infall Times Using Machine Learning}

\author[Barmentloo \& Cautun]
{\parbox{\textwidth}{
    Stan Barmentloo$^{1}$ \thanks{E-mail: s.barmentloo@outlook.com}
    and Marius~Cautun$^{1}$  \thanks{E-mail: cautun@strw.leidenuniv.nl}
    }
\vspace{.2cm}\\
$^{1}$ Leiden Observatory, Leiden University, PO Box 9513, NL-2300 RA Leiden, the Netherlands \\
}


\pubyear{2022}

\label{firstpage}
\pagerange{\pageref{firstpage}--\pageref{lastpage}}
\maketitle

\begin{abstract}
   A key unknown of the Milky Way (MW) satellites is their orbital history, and, in particular, the time they were accreted onto the MW system since it marks the point where they experience a multitude of environmental processes.
   We present a new methodology for determining infall times, namely using a neural network (NN) algorithm. The NN is trained on MW-analogues in the EAGLE hydrodynamical simulation to predict if a dwarf galaxy is at first infall or a backsplash galaxy and to infer its infall time. The resulting NN predicts with 85\% accuracy if a galaxy currently outside the virial radius is a backsplash satellite and determines the infall times with a typical 68\% confidence interval of 4.4 Gyrs. 
   Applying the NN to MW dwarfs with Gaia EDR3 proper motions, we find that all of the dwarfs within 300 kpc had been inside the Galactic halo. The overall MW satellite accretion rate agrees well with the theoretical prediction except for late times when the MW shows a second peak at a lookback time of 1.5 Gyrs corresponding to the infall of the LMC and its satellites. We also find that the quenching times for ultrafaint dwarfs show no significant correlation with infall time and thus supporting the hypothesis that they were quenched during reionisation. In contrast, dwarfs with stellar masses above $10^5\Msun$ are found to be consistent with environmental quenching inside the Galactic halo, with star-formation ceasing on average at $0.5^{+0.9}_{-1.2}$ Gyrs after infall.
\end{abstract}


\section{Introduction} 
\label{sec:introduction}
Within the standard $\Lambda$ Cold Dark Matter ($\Lambda$CDM) cosmological model structures form hierarchically though the merger of lower mass galaxies and haloes. The signature of this process are satellite galaxies that orbit a typically more massive central galaxy and the stellar haloes that surround most galaxies. The former are the remnants of defunct galaxies that either merged with or were tidally destroyed by their hosts (e.g. the \citealt{Frenk2012} and \citealt{Zavala2019} reviews and references within). 

The Milky Way (MW) represents the perfect test bed for studying the hierarchical growth of haloes and galaxies due to its close proximity and a wealth of very detailed observations. In particular, we have a census of nearly 50 Galactic satellites and many thousands of halo stars with full 6D phase-space observations \citep[e.g.][]{Mcconnachie2020propermotions,GaiaDR2,Gaia2021_EDR3}. These have shown that our galaxy experienced two massive early mergers, Gaia-Enceladus-Sausage and Kraken \citep[e.g.][]{Belokurov2018Sausage,Helmi2018Sausage,Kruijssen_e-mosaics_2019}, and the more recent accretion of the LMC around 2 Gyrs ago \citep[e.g.][]{Besla2010LMCrecent,Cautun2019LMC,Patel2020LMCSMC}. This is a first step into revealing the MW's assembly history, with a much more detailed picture emerging when studying the infall times of all Galactic satellites. 

The MW satellites also offer the most detailed observations of dwarf galaxies and contain unique signatures on the nature of dark matter (DM) and galaxy formation processes \citep[e.g. the review of][]{Bullock2017Challenges}. However, to interpret the Galactic satellite data we need to know when these objects were accreted onto the MW halo. For example, this tells us which processes quenched the star formation of the Galactic satellites, with all MW dwarfs except LMC and SMC having no star-forming gas \citep{Putman2021_HI_content}. Currently, there are two competing theories for explaining dwarf galaxy quenching. The reionisation of the Universe is predicted to have removed the gas from low mass galaxies and thus stopped star formation \citep[e.g.][]{Bullock2000Reionisation,Benson2002,Sawala2010,Simon2019UFDreview}. This is expected to be the dominant process for ultra-faint dwarfs (UFD), with most having stopped forming stars 11-13 Gyr ago \citep[e.g.][]{Brown2014Metalpoor,Weisz2014_UFD_quenching,Sacchi_2021}. On the other hand, the more massive classical dwarfs probably keep most of their gas reservoir after the epoch of reionisation and thus continued forming stars. The moment at which they stop forming stars is not so much dependent on global processes (e.g. reionization), but rather on the specific history of the individual dwarf galaxies, such as ram-pressure quenching when they become satellites of a more massive galaxy \citep[e.g.][]{Gatto2013Ramming, Simpson2018,Akins2021_quenching}. The mass threshold separating quenching by reionisation from environmental effects is still debated and it represents a key probe of star-formation processes in the smallest galaxies \citep[e.g.][]{Bose2018Reionisation,Simon2019UFDreview,Benitez-Llambay2020}. 

The Galactic dwarfs are currently one of the most constraining probes into the nature of DM \citep[e.g.][]{Enzi2021_WDM,Nadler2021_WDM,Newton2021_WDM}, and augmenting existing studies with infall time and orbit information can improve the constraints further \citep[e.g.][]{Kaplinghat2019}. Alternative DM models, such as warm DM or self-interacting DM, predict differences in the number and structure of low mass galaxies and halos compared to the standard cold DM model \citep[e.g.][]{Colin2000_WDM,Zavala2013_SIDM,Lovell2014_WDM}. These differences depend on the orbital history of the satellites since dwarfs in alternative DM models, which typically have shallower DM density profiles and later formation times, experience enhanced tidal stripping compared to their cold DM counterparts \citep[e.g.][]{Dooley2016_ADM_stripping,Lovell2021_WDM_vs_CDM}, which emphasises the importance of accurate accretion times for the Galactic satellites. 

Determining the infall times of the MW satellites is a key question in cosmology and has been the subject of many previous studies. These can be grouped in two broad categories. The classical approach is to integrate the orbits of the satellites backwards in time and determine when they first crossed the virial radius of their host \citep[e.g.][]{Besla2010LMCrecent,Cautun2019LMC,Patel2020LMCSMC}. However, this problem is inherently difficult due to the unknown evolution of the Galactic potential and due to the chaotic nature of satellite-satellite interactions \citep[e.g.][]{D'Souza2022}. Studies employing this approach typically involve many simplified assumptions, such as neglecting satellite-satellite interactions and assuming a smooth spherically symmetric and slowly varying MW potential, which makes it difficult to estimate robust uncertainties in the inferred satellite accretion times \citep[e.g.][]{Miyoshi2020Infall1st,Armstrong2021}.

A second approach for determining the infall time is to match the observed phase space distribution of observed satellites with those of satellites in cosmological simulations \citep[e.g.][]{Rocha2012Infall2nd,Fillingham2019Quench}. This has the advantage of capturing the many processes that affect satellite orbits since these are included in the simulation by construction. However, one of the main limitations arises from the technique used to match observed satellites with their simulated analogues, since it is not known which satellite properties are most important for determining the infall time. \citet{Rocha2012Infall2nd} have claimed that binding energy is the main predictor of infall time, however later simulations have found a large scatter in this relation, especially for systems that experienced massive accretions \citep{D'Souza2022}. \citet{Fillingham2019Quench} further improved upon this matching procedure by, on top of the binding energy, matching observations and simulations also in terms of distance and radial velocity. However, this raises a major difficulty since it unclear what the 'closest' means when matching many different physical variables.

\bigskip

In this research, we present a new approach to determine the infall time of satellite galaxies using neural networks. Machine learning is an excellent solution for the problem of determining infall times, as it specialises in searching for potential correlations that can not be easily seen by researches due to being complex and involving multi-dimensional spaces. This new method is similar to matching observed and simulated satellites in phase space with the major advantage that the machine learning algorithm takes care of determining the optimal weights of the different variables internally, avoiding incorrect assumptions. To train the machine learning, we make use of the galaxy data from the EAGLE hydrodynamical simulation \citep{Main_Eagle_Schaye}. EAGLE represents a good compromise between large volume, which is needed to have many MW-analogues, and sufficient resolution to resolve tens of satellites for each MW-mass system. We then calculate the infall time likelihood for 47 Galactic satellites that have 6D phase space data, while accounting for uncertainties in the MW mass model and in the observed properties of satellite galaxies.

The paper is structured as follows. First, the simulation data used for training our machine learning algorithm as well as the observational data for the MW satellites is described in section \ref{chapter:Data}. Next, the workings of our adopted algorithm as well as its capabilities are discussed in section \ref{chapter:Methods}. Our results are given in section \ref{chapter:results}, to be extensively analysed in section \ref{chapter:Discussion}. Finally, section \ref{chapter:conclusions} reiterates the main finds deduced from our results.

\section{Data}
\label{chapter:Data}

Here we describe the data used to train and test the machine learning algorithm (sections \ref{sec:simulations}-\ref{sec:param_selec}) and how we process the data for the MW satellites such that it can be used by our machine learning pipeline (section \ref{sec:obs_data}).

\subsection{Simulations}
\label{sec:simulations}

\subsubsection{EAGLE}
\label{sec:EAGLE:data}

The Evolution and Assembly of GaLaxies and their Environments (EAGLE) project is ``a suite of hydrodynamical simulations that follow the formation of galaxies and supermassive black holes in cosmologically representative volumes of a standard $\Lambda$CDM universe" \citep{Main_Eagle_Schaye}. The simulations include a multitude of processes that are thought to be key for the formation and evolution of galaxies, such as metal enrichment, energy feedback from star formation and the accretion and mergers of supermassive black holes, and have been shown to reproduce many properties of the galaxy population \citep[for the details, see][]{Main_Eagle_Schaye,Crain2015EAGLE}.

All training data for the machine learning in this research is taken from the main EAGLE simulation that is labelled as `Ref-L0100N1504'. This is the largest of the EAGLE project simulations and corresponds to a cube with side-lengths of 100 Mpc that contains an equal number of $1504^{3}$ DM and gas particles of mass $9.6$ and $1.8\times10^6\Msun$ respectively. This simulation is ideal for obtaining a large number of MW-analogues systems and their satellites that can be used to train our machine learning pipeline. 

\subsubsection{Auriga}

We also want to test to what extent our machine learning predictions are sensitive to the use of one specific simulation. For example, artefacts could arise from the use of one specific galaxy formation model as well as from the rather limited numerical resolution with which satellite galaxies are resolved in EAGLE \citep[e.g.][]{Lacey1993,Qi2010,vandenBosch2018a}. As such, we make use of a second suite of 30 MW-mass zoom-in hydrodynamical simulations that have been run as part of the Auriga project \citep{Auriga}. These simulations employ a different galaxy formation model that is similar to that used in the Illustris-TNG \citep{Pillepich2018} and have a 30 times better mass resolution than the main EAGLE run \citep[for more details see][]{Auriga}. Due to its size, the Auriga data is too small to properly train a neural network. As such, we will use the Auriga satellite galaxies to test the accuracy of our machine learning method that has been trained only on the EAGLE data. 

\subsection{Sample Selection}
\label{subsec:sample_selection}

The dataset used to train the machine learning algorithm serves as a model prior for the MW and thus we should select systems that best resemble our galaxy and its environments.
We define a MW analogue as a system whose total mass is comparable to that of our own galaxy, which is around $10^{12} M_{\odot}$ \citep[e.g.][]{Cautun2020MW,Wang2020}. Furthermore, the dynamics of the satellites should be dominated by one large galaxy, in our case the MW. The closest massive neighbour to the MW with more than half of its mass is the Andromeda galaxy at about 770 kpc (\citealt{Mcconnachie2012SatCoords}). This corresponds to a distance of roughly 3.5 R$_{200}$ from our galaxy, where R$_{200}$ is defined as the distance from the galaxy centre where the average enclosed density is 200 times the critical density. Most studies refer to Galactic satellites as all the galaxies within 300 kpc from the Galactic Centre \citep[e.g.][]{Bullock2017Challenges,Shao2019}, which corresponds to a distance of ${\sim}1.5 R_{200}$.

Other properties of the MW are also thought to affect the infall times of its satellites, such as the accretion of massive satellites, halo growth history and local environment \citep[e.g.][]{Fakhouri2009,Deason2015,Bose2020,dsouza2021massive}. Including one or more of these criteria can result in closer MW-analogues, however, we choose not to due so since we do not want to be overly restrictive in our sample selection. This is motivated by the goal of having a large training sample, of testing the predictions against higher resolution simulations that contain only a small number of MW-mass hosts, and of not imposing our own potentially incorrect biases. Nevertheless, increasing the number of MW selection criteria can reduce halo-to-halo scatter and could lead to a more accurate measurement of satellite infall times.

More specifically, the following two criteria were used to select present-day MW-analogues:
\begin{enumerate}
    \item The host galaxy has a mass $M_{200}$\footnote{The mass contained in a sphere of radius $R_{200}$, the radius at which the average density is equal to 200 times the critical density.} in the range $[0.5,2.0]\times10^{12}$.
    \item The host galaxy has no massive neighbour, that is another galaxy within $2 R_{200}$ whose total mass is larger than $0.5 M_{200}$.
\end{enumerate}

Our satellite sample consists of all subhalos found within a distance of $2 R_{200}$ from the centre of the host galaxy. We include all subhalos, not only luminous ones (i.e. with stars), since due to the limited resolution of the EAGLE simulation the lowest stellar mass of an object is $10^6\Msun$. However, some MW satellites have stellar masses as low as ${\sim}10^3 \Msun$. Many of the subhalos hosting such faint galaxies are resolved as dark matter only substructures in EAGLE, which is why we consider all subhalos when finding satellites. Furthermore, normally satellites are taken as the galaxies within $R_{200}$, however many so-called Galactic satellites are found at larger distance (see discussion above) and thus we choose a larger radius to identify satellites. Even if some galaxies are found outside $R_{200}$ at present day, they could have been inside the virial radius of the host at earlier times (so-called backsplash galaxies; e.g. \citealt{Wetzel2014,Simpson2018}).

These selection criteria have resulted in 1,628 present-day MW-analogues that contain a total 70,468 satellites above the resolution limit of the main EAGLE run. To simplify the calculation of the infall time (see section \ref{subsec:infal_time_calculation}), we further removed all galaxies, both centrals and satellites, that since formation have crossed an edge of the simulation box (i.e. if one of their positional coordinates jumped from $\sim$ 100 Mpc to $\sim$ 0 Mpc). This left a final sample of 1,590 hosts and 63,402 satellites. 

Once all present-day galaxies were selected, they were traced back in time using the galaxy merger-tree available on the EAGLE public database \citet{Mergertree_Eagle_McAlpine}. This consists of the most massive progenitor branch of the merger tree. We stored the data for all snapshots in which both the central and the satellite galaxy exists (EAGLE sometimes loses track of a galaxy in a snapshot, for it to reappear in the following ones).

We further limit the satellite population to the ones that are expected to host the majority of luminous galaxies. A multitude of galaxy formation prescriptions, such as the semi-analytical models \citep[e.g.][]{Wang2013}, very high resolution hydrodynamical simulations \citep[e.g.][]{Sawala2016b,Wheeler2019,Applebaum2021,Grand2021} and theoretical models on how gas cools and fragments to form stars \citep{Benitez-Llambay2020}, suggest that most galaxies form in haloes of total mass larger than ${\sim}10^{9}~\Msun$. We implement this selection as a threshold on the peak maximum circular velocity\footnote{We determine $V_{\rm peak}$ as the peak of the maximum circular velocity, $V_{\rm max}$, for the most massive progenitor branch. For satellites, $V_{\rm peak}$ is typically given by the value of $V_{\rm max}$ just before infall onto the host halo.}, which we denote as $V_{\rm peak}$, since the stellar mass has been shown to have a tighter correlation with $V_{\rm peak}$ than with halo mass \citep[e.g.][]{Matthee2017,Fattahi2018,Garrison-Kimmel2019}. 

We define our satellite sample as the subhaloes with $V_{\rm peak} \geq 25 ~\rm{km/s}$. This is motivated two-fold. Firstly, as we just discussed, most galaxies form in massive haloes and the $V_{\rm peak} = 25 ~\rm{km/s}$ corresponds to the sweet spot where we expect that around half of those haloes to contain a galaxy \citep{Sawala2016b,Ethan2021LMCQuenching}. Secondly, low mass satellites are close to the resolution limit of the simulation and their internal structure is not well resolved, which can introduce numerical artefacts in their orbital evolution, such as premature tidal disruption \citep[e.g. see][]{vandenBosch2018a,Grand2021}. These numerical artefacts would preferentially affect early accreted substructures, since these spend more time within the virial radius of the host, and thus could bias the distribution of infall times. Based on the \citet[][see also \citealt{Hellwing2016}]{Millenium2010} analysis of halo structure in the Millennium II simulations, whose mass resolution is very close to that of the main \eagle{} simulation, \eagle{} resolves robustly only halos with $V_{\rm peak} \geq 25 ~\rm{km/s}$.
When imposing this $V_{\rm peak}$ selection, which affects only the satellite sample, we are left with 30515 satellites. 

We have used the same exact selection criteria also for the \auriga{} sample. All 30 \auriga{} systems pass our MW-analogue selection since all of them were chosen to be isolated halos with total masses in the range, $M_{200}\in[1.0,2.0]\times10^{12}\Msun$ \citep[see][]{Auriga}. The satellite sample consisted of all subhalos within a distance of $2R_{200}$ of each MW-analogue that have a peak maximum velocity, $V_{\rm peak} \geq 25 ~\rm{km/s}$.

\subsubsection{Infall Time Determination}
\label{subsec:infal_time_calculation}

We define the infall time to be the moment at which a satellite for the first time crosses the virial radius, $R_{200}$, of its present-day MW-mass host. Many of these satellites were isolated dwarfs before accretion onto their $z=0$ hosts \ \citep{Shao2018a}. However, some of them would have been accreted as part of a group, that is they are so-called satellites-of-satellites \citep{Deason2015,Wetzel2015,Ethan2021LMCQuenching}. We do not distinguish between the two, except when discussing in section \ref{sec:MCdisc} this aspect in relation to the satellites brought in the MW by the LMC.

\begin{figure}
    \centering
    \vspace{-.4cm}
    \includegraphics[width = 1.1\linewidth]{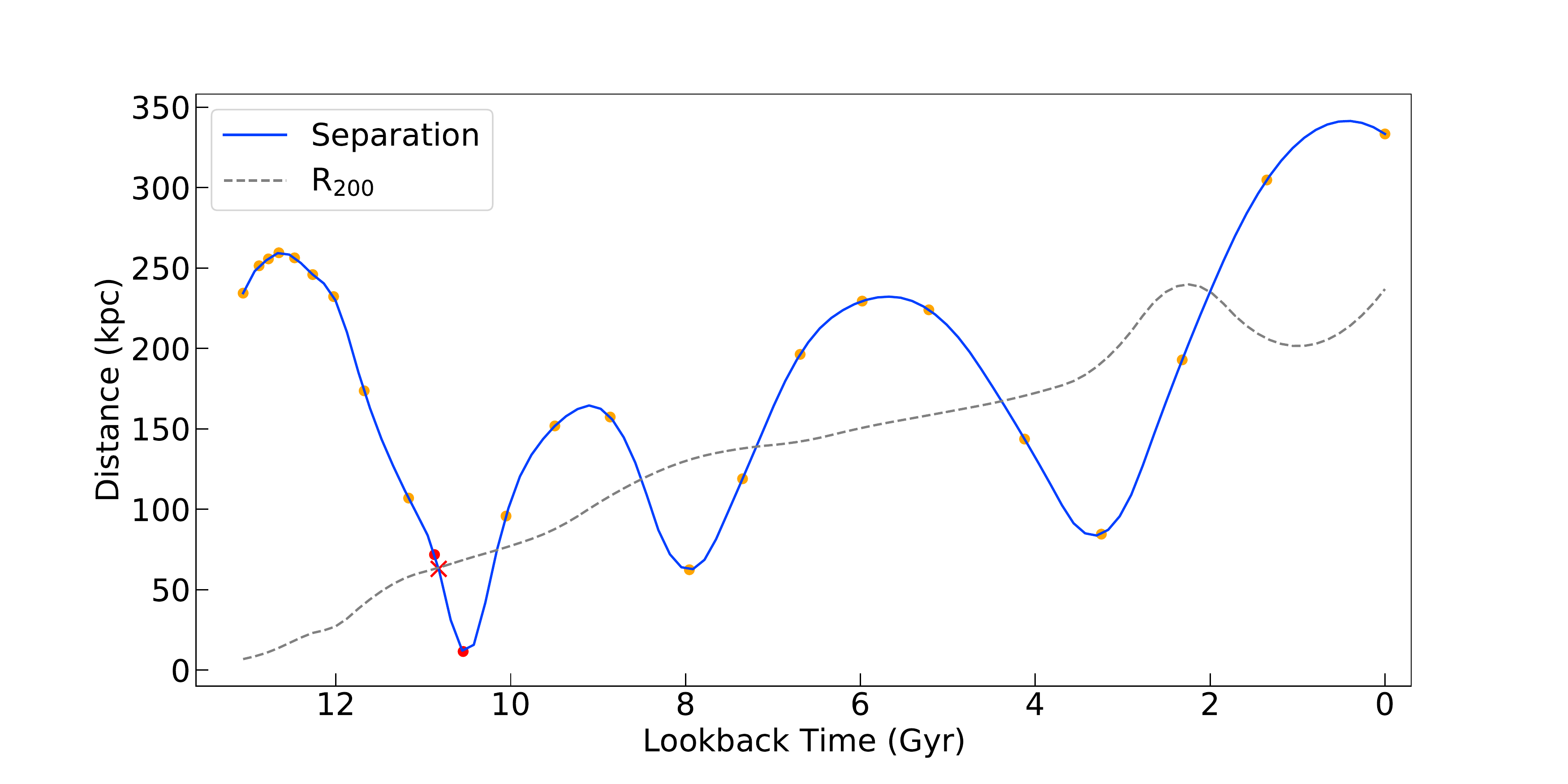}
    \vspace{-.4cm}
    \caption{Example of an orbit of a satellite around a MW-mass host. It shows the proper separation between satellite and host as a function of time. This is measured at multiple snapshots, shown as orange dots, and, for clarity, the blue line shows a spline interpolation between these points. The dotted line shows the radius $R_{200}$ of the host, which again is a spline interpolation between snapshots. The infall time is defined as the first time the satellite enters the virial radius of its present-day host and it is indicated on the figure by a red cross.}
    \label{fig:splashback}
\end{figure}

As the simulation has a discrete number of snapshots (with a typical time interval between snapshots of $\sim$ 0.2 Gyr at early times and $\sim$1 Gyr at late times), determining the exact infall time requires interpolation between snapshots. As we will discuss in section \ref{chapter:results}, the typical uncertainty on the inferred infall time is much higher than the time difference between snapshots, which makes linear interpolation sufficient. That is, the infall time is given by:
\begin{equation}
    t_{\rm infall} = t_{1} +\frac{ \frac{r_{1}-R_{200, 1}}{R_{200, 1}} }{ \frac{r_{1}}{R_{200, 1}} - \frac{r_{2}}{R_{200, 2}} } (t_{2}-t_{1})
    \label{eq:1}
\end{equation}
with t$_{1}$, t$_{2}$ the time of the snapshot before and after infall, r$_{i}$ the comoving distance at time t$_{i}$ of the satellite with respect to the central galaxy, and R$_{200, i}$ the radius of the host at t$_{i}$. This procedure is illustrated in Figure \ref{fig:splashback}, which shows the orbit of a random satellite. The figure also shows that satellites can go in and out of the virial radius of their host multiple times (while outside $R_{200}$ they are referred to as backsplash objects) and also that host $R_{200}$ does not always increase smoothly with time, with rapid increases and decreases taking place when another massive halo is accreted or flies by (e.g. see bump in $R_{200}$ at a lookback time of 2 Gyrs).  

\begin{figure}
    \centering
    \vspace{-.7cm}
    \includegraphics[width = 1.0\linewidth, height = 0.7\linewidth]{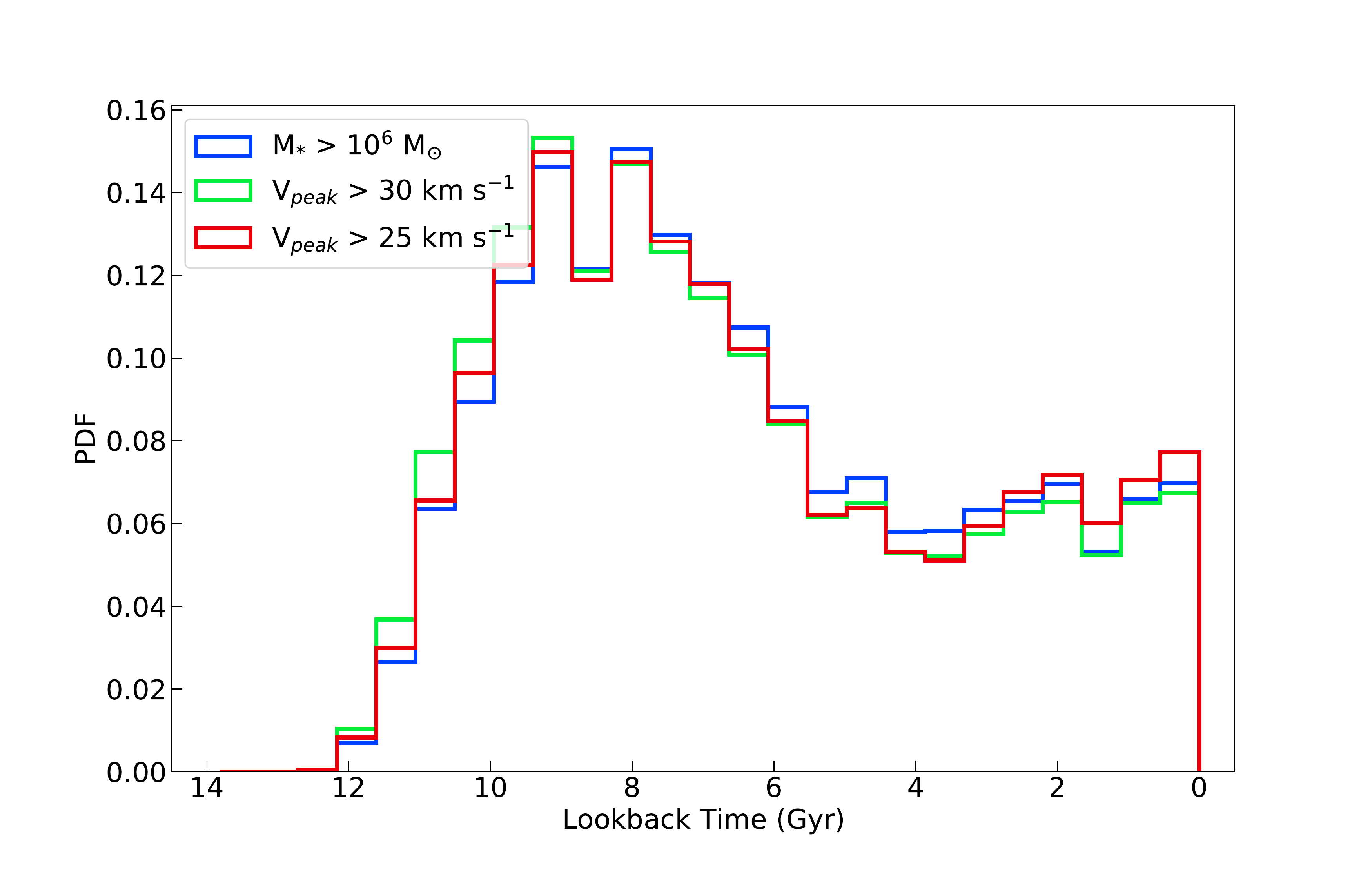}
    \vspace{-.7cm}
    \caption{Distribution of the infall times for satellites of MW-mass hosts in the EAGLE simulation. We present results for three samples: galaxies with stellar mass above $10^6\Msun$, and subhaloes with peak maximum circular velocity, $V_{\rm peak} > 30$ and $25 ~\rm{km/s}$. All samples have roughly the same distribution of infall times. The PDFs include all satellites found within a distance of $2R_{200}$ from the host halo at $z=0$.}
    \label{fig:VMax_vs_tinfall}
\end{figure}

The resulting infall time distribution is shown in Figure \ref{fig:VMax_vs_tinfall}, with the fiducial sample results being shown by the red curve. We find that the infall time probability distribution function (PDF) has a pronounced peak at a lookback time of $9~\rm{Gyrs}$ ago, a sharp cut-off at earlier times and a more gradual decreases towards present day, being roughly flat for the last $6~\rm{Gyrs}$. The oldest surviving satellites of our MW-analogues have been accreted $12~ \rm{Gyrs}$ ago. The infall time distribution is shaped by two competing effects, which are clearly illustrated in \citet[see their Figures 2 to 4]{Fattahi2020}. Firstly, the satellite accretion rate is largest at early times when the universe was smaller and when halos grow very fast, typically through mergers. Secondly, the survival rate of satellites is inversely correlated with the time they orbit inside their host. The more time they spend as satellites (i.e. the earlier they were accreted) the lower is their chance to survive to present day.

\begin{figure*}
    \centering
    \includegraphics[width = .85\linewidth]{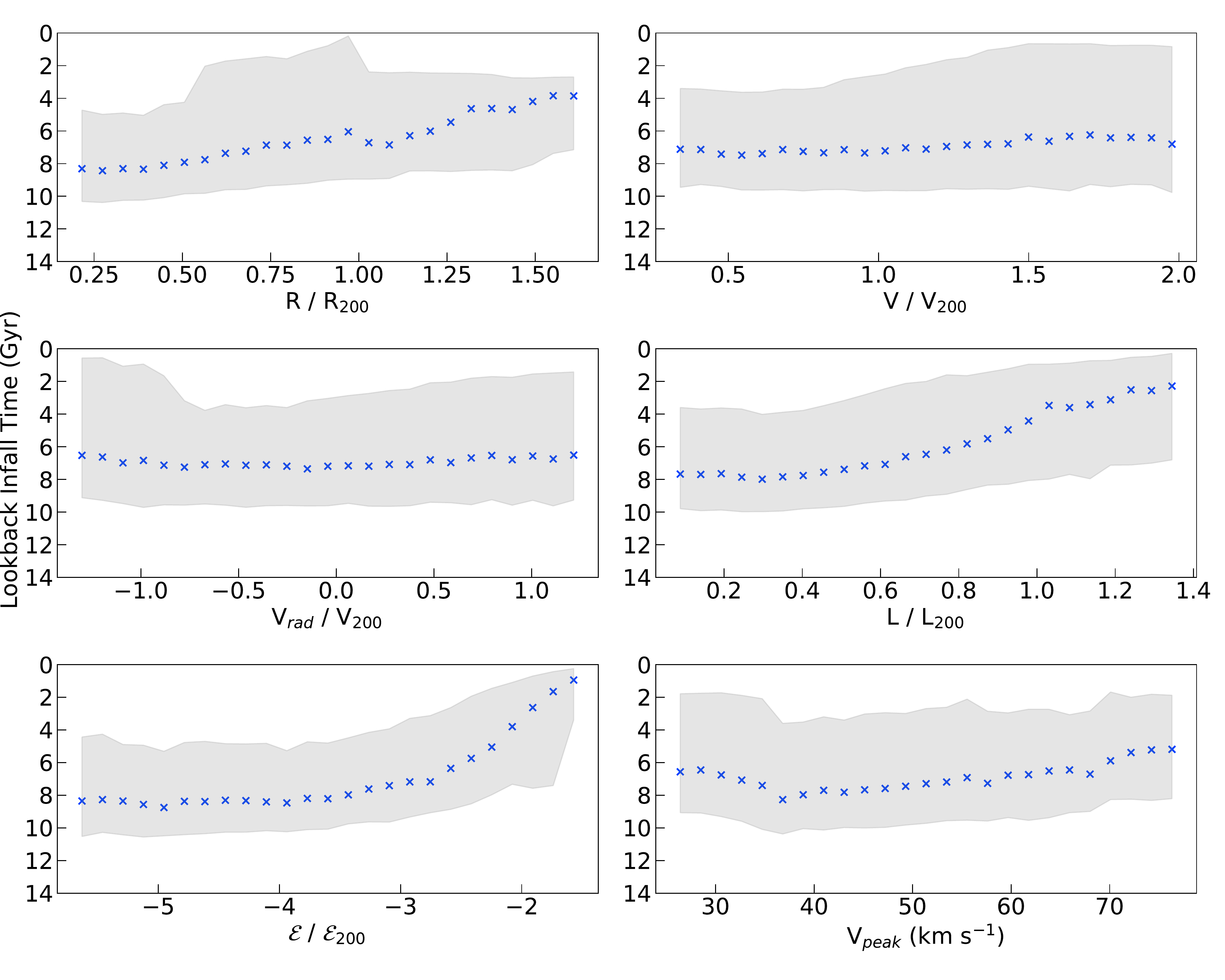}
    \vskip -.2cm
    \caption{Infall times as a function of the five input features used to train the machine learning algorithm (the bottom-right quantity, $V_{\rm peak}$, was not used as input and is shown here only for illustrative purposes). The input features are as follows: distance, velocity magnitude, radial velocity, specific angular momentum, and specific orbital energy, and are scaled by the host properties to make them invariant to the mass of their host (see main text for details). The blue crosses show the median infall times for each bin and the grey areas show the 16 to 84 percentiles in each bin. }
    \label{fig:tinfall_vs_binned_Parameter_values}
\end{figure*}

As we have discussed in section \ref{subsec:sample_selection}, our satellite population is selected as the objects with $V_{\rm peak} \ge 25~\rm{km/s}$. In Figure \ref{fig:VMax_vs_tinfall} we also investigate if this selection biases the infall time distribution compared to a stellar mass selection or to using another $V_{\rm peak}$ threshold. We find that the infall time PDF is approximately the same for all three selections. This was to be expected since previous studies have shown that the distribution of infall times for satellites of MW-mass hosts is largely independent of their stellar mass except for the most massive objects with $M_\star \geq 10^8~\Msun$ \citep[e.g.][see also bottom-right panel in Figure \ref{fig:tinfall_vs_binned_Parameter_values}]{Shao2018b,Fattahi2020}. For the massive satellites, due to their high total mass, dynamical friction plays an important role and thus high stellar mass satellites have typically more recent accretion times.

\subsection{Feature Selection}
\label{sec:param_selec}

Our goal is to estimate the infall time using the orbital phase space information of satellite galaxies that has recently become available for a large number of MW dwarfs \citep[e.g.][]{Mcconnachie2020propermotions,Battaglia2021_EDR3_pm}. As such, to train our machine learning framework we will use the 3D position and velocity of the satellite with respect to the host centre, which we summarise in terms of: i) distance from the host, ii) total velocity magnitude, iii) radial velocity component, and iv) specific angular momentum. 

Based on earlier studies, e.g. \citet{Rocha2012Infall2nd}, we expect that there is a strong correlation between satellite orbital energy and infall time. The specific energy of a satellite is the fifth feature used as input to our machine learning method. This is the sum of the relative kinetic energy per unit mass and the gravitational potential of the satellite. To calculate the latter, we need the mass profile of the host. For the EAGLE sample, we make use of the four-component fit to the total density profile of EAGLE galaxies introduced by \citet{Halo_Energies_Schaller} and calculate the gravitational potential at the position of each satellite using equation (21) in that paper. The \citeauthor{Halo_Energies_Schaller} functional form has a greater flexibility than the typical Navarro, Frenk and White (NFW; \citealt{NFW1996}) profile used to described halos in dark matter only simulations and gives a much better fit to the total density profile in hydrodynamical simulations that include, beside a dark halo, a central stellar component and an extended hot gas distribution.

We have studied using other galaxy observables as input for infall time determination,
however we decided against including them in our fiducial model. One such feature is a satellite's stellar mass since, for example, more massive satellites needed longer to grow and thus would be accreted on average later. Due to the limited resolution of the \eagle{} simulation, whose star particles have a mass of ${\sim}10^{6}$ \Msun, it would mean that either the majority of satellites would have a missing stellar mass value or, if we would have limited the study to satellites with well determined stellar masses (e.g. at least 10 stellar particles), would severely reduce our training sample and its applicability to the MW satellites. Similar reasons motivated not using other observables such as galaxy colours.

\subsubsection{Parameter Scaling}

The satellites in EAGLE orbit a wide range of host galaxies. Satellite observables such as position and velocity depend on the size and mass distribution of their host \citep{Callingham2019Scaling,RodriguezWimberly2021}. This dependence can be almost fully removed by scaling with the properties of the host galaxy, since the satellite systems have a similar structure across a wide range of host halo masses \citep{Callingham2019Scaling,Li2017,Li2020}.

To train the machine learning algorithm we use scaled satellite properties since this way we can eliminate the host mass as one of the input features.
Distances and positions were scaled using the host radius, that is we use the quantity $r/R_{200}$, where $r$ is the satellite distance from the host. We scaled the velocity magnitude and its components using the circular velocity at the halo radius, $V_{200}=\sqrt{\tfrac{G M_{200}}{R_{200}}}$. The specific angular momentum was scaled by the angular momentum an object on a circular orbit with velocity V$_{200}$ at a distance R$_{200}$. 
Finally, the energy was scaled by the specific kinetic energy of an object on a circular orbit at R$_{200}$, whose value equals $\frac{1}{2}$V$_{200}^{2}$.

The correlation between each machine learning input feature and the satellite infall time is shown in Figure \ref{fig:tinfall_vs_binned_Parameter_values}. To better illustrate these trends, we split the data in bins of the property shown along the x-axis and we show the median and the 16 to 84 percentiles of the $t_{\rm infall}$ distribution in that bin. Two results are made clear by the plots in Figure \ref{fig:tinfall_vs_binned_Parameter_values}.

Firstly, the infall time depends more strongly on some parameters than on others. For example, the infall time shows the strongest dependence on energy (as was previously observed in, for example, \citealt{Rocha2012Infall2nd}), especially when reaching to higher, less bound energies. At the same time, the two velocity features show almost no correlation with infall time. One should keep in mind however that when considering the dependence on multiple parameters simultaneously, some of the now flat correlations might turn out to play an important role. 

Secondly, the distribution of infall times for fixed values of any of the five input features is rather wide. Even for the specific energy, the width of the conditional $t_{\rm infall}$ PDF is at least $6~\rm{Gyrs}$ or larger. This emphasises the complexity of determining infall times from current day observables and gives an indication of the typical uncertainties which should be expected for the machine learning prediction. 

\subsubsection{The impact of numerical resolution on subhalo infall time}
In the bottom-right panel of Figure \ref{fig:tinfall_vs_binned_Parameter_values} we see that the median \tinf{} decreases slowly with decreasing subhalo peak circular velocity until around $V_{\rm peak}=35\kms$, after which the trend reverses. This non-monotonic relation indicates that the infall times of subhalos with $V_{\rm peak}<35\kms$ are likely affected by the finite resolution of the simulation and that this threshold is somewhat larger than the 25\kms{} value found when analysing the convergence of subhalo internal properties \citep[e.g.][]{Millenium2010}.

This numerically-induced artificial tidal disruption becomes more important the smaller the number of particles with which a subhalo is resolved. This raises the question of determining an optimal $V_{\rm peak}$ threshold such that we include as many subhalos as possible while mitigating any biases arising from artificial disruption. For our study, the quantity of interest is the \tinf{} PDF of all the subhalos that will be used to train the neural network since any bias in the PDF will result in biased predictions. This is why we have studied how the \tinf{} PDF changes as we include lower mass (i.e. lower $V_{\rm peak}$) subhaloes in our sample. As we decrease the $V_{\rm peak}$ threshold and include ever more subhalos, we find that the \tinf{} PDF is largely insensitive to the value of the threshold as long as we limit to $V_{\rm peak} > 25\kms$. For example, this is illustrated in Figure \ref{fig:VMax_vs_tinfall}, which shows only small differences between the \tinf{} PDF of subhalos with $V_{\rm peak} > 25\kms$ and of those with $V_{\rm peak} > 30\kms$. Once we decrease this threshold further (not shown here), we find a rapid change in the \tinf{} PDF indicating that artificial subhalo disruption plays a very important role.

The fact that the \tinf{} PDF does not change strongly with subhalo $V_{\rm peak}$ is the outcome of two competing effects. As we study smaller subhalos, we expect a small preference towards earlier infall times. Once we analyse small enough subhalos, numerical effects kick in and lead to an opposite trend: a tendency for later infall times. For the \eagle{} simulation, the two effects balance each other at $V_{\rm peak}{\sim}30\kms$. This means that choosing a slightly lower $V_{\rm peak}$ selection limit, let us say $25\kms$, leads to equally accurate results as a higher value, e.g. $35\kms$, but has the added benefit of including a considerable larger population of subhalos.

\subsection{Observational Data}
\label{sec:obs_data}

For 47 the MW satellites with 6D phase space data, we adopt the distance, position and radial velocity from \citet{Mcconnachie2012SatCoords}, which is a compilation of various measurements of nearby dwarf galaxies. The proper motions are taken from \citet{McConnachie2020b}, where Gaia EDR3 proper motions for individual stars were combined with a photometric and radial velocity analysis. That study, which is based on the method introduced in \citet{Mcconnachie2020propermotions}, has used a Bayesian formalism to identify likely dwarf galaxy member stars that were than used to calculate the average proper motions of each dwarf. In general, the inferred proper motions are in good agreement with previous results using HST, {\it Gaia} DR2, and studies using spectroscopically confirmed dwarf member stars, but have smaller uncertainties \citep[see comparison in][]{Battaglia2021_EDR3_pm}. 

To account for measurement errors in Galactic satellite properties, we use Monte Carlo sampling. For each satellite, we created 1000 random samples in observed space, i.e. distance modulus, radial velocity, and proper motions, assuming that the measurements are described by a Gaussian distribution whose centre is the measured value and whose width is given by the measurement errors. These values are then transformed to a Galactocentric reference frame  using the following values for the solar position and motion: Sun's distance from the Galactic Centre, $R_\odot= (8.178 \pm 0.022) ~\rm{kpc}$ \citep{Gravity-2019}, circular velocity at the Sun's position, $V_{\rm circ} = (234.7\pm1.7)~\rm{km/s}$ \citep{Nitschai-2021}, and the Sun's motion with respect to the local standard of rest, $(U,V,W)=(11.10 \pm 0.72, 12.24\pm0.47, 7.25\pm0.37)~\rm{km/s}$ \citep{Schonrich-2010}.

To determine the specific energy of each Galactic satellite, we model the MW systems as a central stellar component and an NFW dark matter halo \citep[e.g. see][]{Cautun2019LMC}. For simplicity, the stellar component was modelled as a point mass distribution, as most satellites are far enough from the stellar disc such that this approximation is justified. The potential of the NFW dark matter halo was taken as
\begin{equation}
    \Phi_{\text {halo }}=-\frac{G M_{200}^{\rm DM}}{r} \frac{\log \left(1+c r / R_{200}\right)}{\log (1+c)-c /(1+c)}
    \; ,
\end{equation}
with $c$ the concentration parameter, $r$ the distance from the Galactic Centre, $M_{200}^{\rm DM}$ the DM halo mass and $R_{200}$ the halo radius of the MW. These values have been taken from the \citet{Cautun2020MW} study, with $M_{200}^{\rm DM} = 0.97^{+0.24}_{-0.19} \times 10^{12}~\Msun$, $c = 9.4^{+1.9}_{-2.6}$, and stellar mass $M_\star = 6.24^{+0.43}_{-0.52} \times 10^{10}~\Msun$. To account for uncertainties in the Galactic mass profile, we generated 1000 Monte Carlo realisations of the total mass and concentration assuming log-normal distributions for these quantities whose mean and width are given by the values quoted above. Finally, we calculate the five machine learning input features and scale them using the procedure described in section \ref{sec:param_selec}.

\section{Methods}
\label{chapter:Methods}

In this work, we are interested to predict the infall time likelihood function for each satellite using the minimum number of assumptions on the shape of the likelihood (as it will become obvious later, in many cases the likelihood is multi-peaked and is not easily modelled as a simple function such as a Gaussian). To achieve this goal we split the range of infall times into multiple bins and cast the inference problem into estimating what is the probability that any given satellite was accreted in the time interval corresponding to a $t_{\rm infall}$ bin. This now becomes a multi-label classification problem, with the number of classes determined by the number of $t_{\rm infall}$ bins.

To predict the infall times we use the Multi-Layer Perceptron (\citealt{RosenBlatt1958}; MLP from here on), which is one type of deep neural network (NN). Typical use cases for MLP algorithms are binary- and multi-label classification tasks, such as fraud detection or determining hand written digits (for further details and use cases, we refer the reader to \citealt{MLP1994} and \citealt{MLP2019}).

We have chosen fifty equidistant infall time bins that span the time interval from the Big Bang up to today. Each bin corresponds to a time interval slightly less than $0.3~\rm{Gyrs}$. The bin width was chosen such that it is considerably smaller than the typical uncertainties in the predicted infall times (as will be shown later in this section), yet large enough to avoid multiple bins containing very few satellites (which would lead to inadequate statistics). We have tested and doubling the number of bins (i.e. halving the bin width) does not affect our predictions, however increasing the number of bins much more does lead to higher uncertainties in our predictions.

The goal is to apply the $t_{\rm infall}$ inference method to all the MW satellites within $300~\rm{kpc}$ from the Galactic Centre. Some of these satellites are outside the MW's radius, $R_{200}$, whose value is ${\sim}220~\rm{kpc}$, which raises the question if such dwarfs are at first infall (i.e. are yet to enter the MW's $R_{200}$ radius) or are backsplash galaxies (i.e. have been inside $R_{200}$ at a previous time). To answer this question, we build another MLP network whose task is to predict the {\it accretion status} of a galaxy outside $R_{200}$, that is the likelihood that a satellite is at first infall or a backsplash galaxy. The technical details of this network as well as the one used to infer $t_{\rm infall}$ are specified in appendix \ref{appendix:NN_architecture}. 

In the rest of this section we investigate the performance of the two NNs on the EAGLE test sample (we used a 60:20:20 split for the training, evaluation, and test steps) and on the Auriga sample of satellites.

\subsection{Performance of the `accretion status' NN}
\label{sec:AON_performance}

\begin{table}
    \centering
    \begin{tabular}{l c  c }
    \hline
    Sample & Completeness &  Purity  \\ 
 & [\%] &  [\%] \\
 \hline \hline 
All satellites   & 85.3 & 86.1 \\
Backsplash         & 91.1 & 88.7 \\
First infall     & 79.5 & 83.4 \\ \hline
    \end{tabular}%
    \caption{Quantifying the success of the NN to classify galaxies outside the host's virial radius, R$_{200}$, into backsplash or first infall galaxies. The results are for the \eagle{} test sample.}
    \label{tab:first_infall_table}
\end{table}

Determining the accretion status (i.e. first infall versus backsplash) of a dwarf  whose present day distance from the central galaxy is larger than R$_{200}$ is a binary classification problem. For each dwarf, the NN outputs the likelihood that the galaxy is at first infall or is a backsplash object. We then assign to that dwarf the label with the largest likelihood, and measure the performance of the NN by calculating the \textit{completeness} and \textit{purity} measures for our prediction. The completeness is defined as the percentage of satellites with true label \textit{A} that are identified by the NN as having label \textit{A}. The purity is the percentage of all dwarfs predicted as having label \textit{A} whose true label is also \textit{A}. 

The completeness and purity of the accretion status NN are given in table \ref{tab:first_infall_table}. We find rather large values, ${\sim}80\%$ or higher, for both quantities indicating that the network is rather successful in distinguishing between backsplash and first infall dwarfs. Both the completeness and purity show that the NN is slightly better at identifying backsplash satellites, which can be partly explained by the fact that the training sample of satellites that are outside their host radius is split 65\%-35\% between backsplash and first infall galaxies, so that most of the training data consists of previously accreted satellites. 

\subsection{Performance of the `infall time' NN}
\label{sec:performance_AT}

The goal of this NN is to estimate the $t_{\rm infall}$ likelihood for each satellite. It does so by estimating the likelihood in 50 $t_{\rm infall}$ equal bins, and, from this, we construct a PDF by linearly interpolating the values of the likelihood which we take to be defined at the middle of each bin. The resulting likelihoods can have a small
bin-to-bin noise associated to them since ultimately only a rather limited subset of the training sample is used to estimate the infall time likelihood for a given satellite (this subset can be thought of as the training samples that the NN estimates as being close in phase-space to the target satellite). We removed this bin-to-bin variation by further smoothing the likelihood using a Gaussian kernel with dispersion equal to twice the $t_{\rm infall}$ bin width, i.e. ${\sim}0.6~\rm{Gyrs}$. As we discuss shortly, this kernel width is considerably smaller than the typical confidence intervals and thus does not have a large impact on the inferred errors.

We determine the infall time as the maximum likelihood value and calculate confidence intervals (CI) using the \citet{Fillingham2019Quench} procedure, namely by integrating the likelihood curve (starting at the maximum likelihood value) until the area it covers equal 68\%. Averaging over all satellites in the \eagle{} test sample, our NN predicts the infall time with a 68\% CI of size $4.4~\rm{Gyrs}$ (i.e. if the CI had been symmetric, this would correspond to standard deviation $\sigma = 2.2~\rm{Gyrs}$). The obtained uncertainties are more than one order of magnitude larger than the bin width use to estimate the $t_{\rm infall}$ likelihood, whose value is $0.3~\rm{Gyrs}$, and thus our choice of bin widths is small enough to not significantly impact the inferred likelihood. The $t_{\rm infall}$ uncertainties represent a considerable fraction of the age of the Universe of ${\sim}13.8~\rm{Gyrs}$ and this further indicates the difficulty of estimating accurate infall times of satellite galaxies.

Our NN-based inference technique uses five features to determine a satellite's infall time and raises the question of how it fares against simpler methods, such as those based on single satellite properties. To keep the comparison simple, we have applied the same exact framework with the exception that we have trained the MLP using only one single feature at a time as input. We have found that the MLP using the specific energy performs the best of all the single-feature approaches, which should not be surprising given the discussion in sections \ref{sec:introduction} and \ref{sec:param_selec}. The specific energy can determine infall times with a 68\% CI of width $5.8~\rm{Gyrs}$, which is considerably larger than the $4.4~\rm{Gyrs}$ uncertainty when using the full set of five features. Significantly larger uncertainties are found when employing the MLPs that use just one of our other four features as their single input. We also have found that applying the MLP to three input features that combine the specific energy and distance with a velocity-based feature (e.g. total or radial velocity) gives equally accurate predictions as our full five-feature model. This is to be expected since many of the input features are correlated and thus contain redundant information.

\subsubsection{Robustness of predictions}
\begin{table}
    \centering
    \begin{tabular}{l c  c }
    \hline
    Sample & $\tinf^{\rm{true}}$ within 68\% CI &  $\tinf^{\rm{true}}$ within 95\% CI  \\ 
 & [\%] &  [\%] \\
 \hline \hline 
\eagle{}   & 69.0 & 94.7  \\
\auriga{} all & 57.7 & 89.0 \\ 
\auriga{} $r>0.75R_{200}$ & 66.6 & 94.4 \\ \hline
    \end{tabular}
    \caption{Testing the accuracy of the NN for predicting the infall time, \tinf, for satellites in the \eagle{} test sample and in the \auriga{} suite of MW-mass simulations. The two columns gives the fraction of satellites whose true infall times, as measured in the simulations, are respectively within the 68 and 95\% confidence interval (CI) of the infall time inferred by the NN architecture. For \auriga{}, we show results for all satellites for for those found at distances larger than $0.75R_{200}$ from their host. The discrepancy between the \eagle{} and \auriga{} results are due to differences in their mass profiles (see discussion in the main text). These results indicate that the NN estimates are realistic CI for \tinf{}.
    }
    \label{tab:test_infall_time}
\end{table}

We checked the robustness of our confidence intervals by determining the fraction of the data for which the 68\% and 95\% CI contain the true infall time, $t_{\rm infall}^{\rm true}$. Ideally these percentages should be around 68\% and 95\%. The results are given in table \ref{tab:test_infall_time}, and for the \eagle{} test sample we find very similar fractions to the expected ones, making us confident in our estimated uncertainties.

\begin{figure}
    \centering
    \vskip -.5cm
    \includegraphics[width = 1\linewidth, height = 1\linewidth]{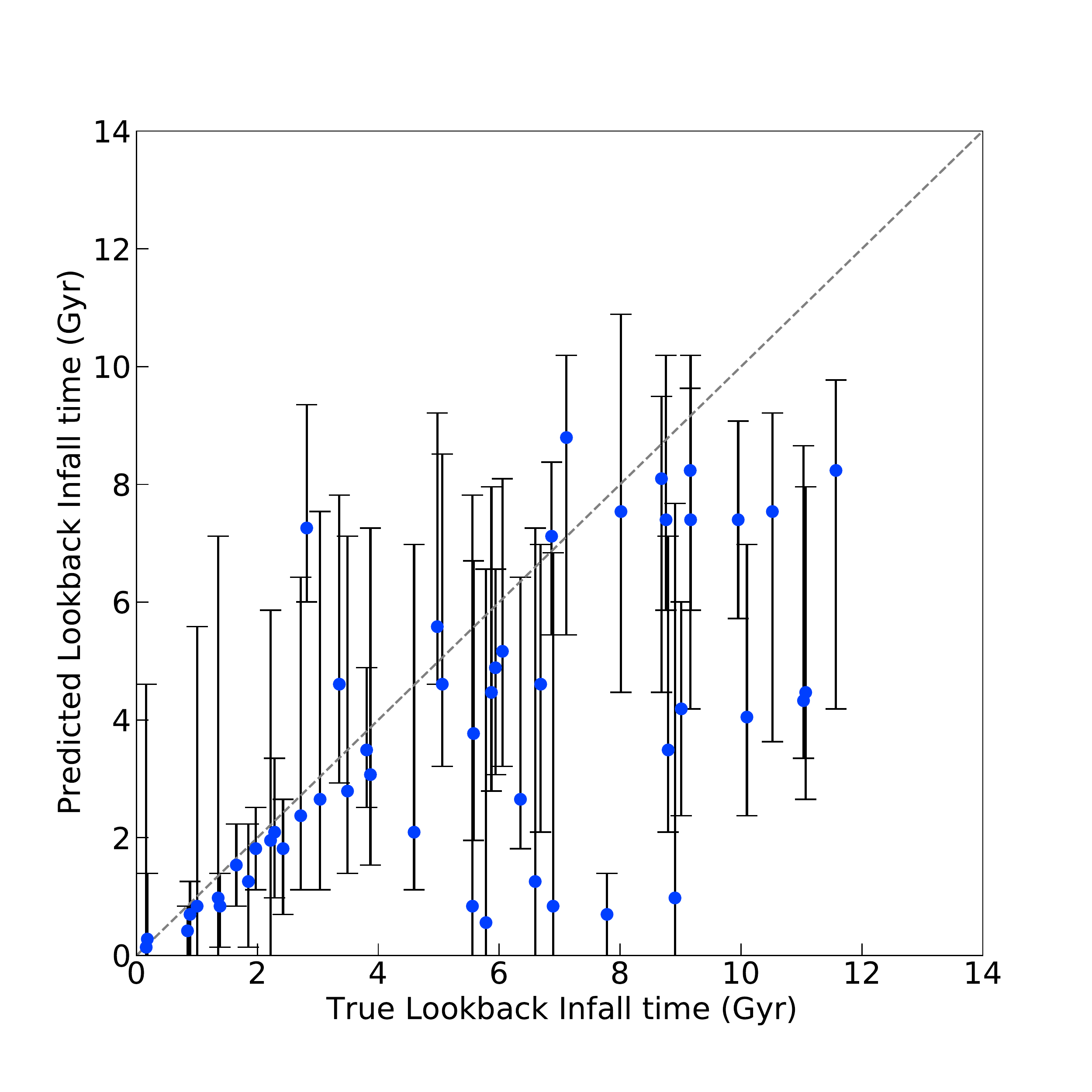}
    \vskip -.3cm
    \caption{
    Comparison between the infall times predicted by the NN, $t_{\rm infall}^{\rm NN}$, and
    the distribution of true infall times, $t_{\rm infall}^{\rm true}$, for 50 random samples from the Auriga sample. The blue filled circles show the maximum likelihood estimates and the errorbars show our inferred 68\% confidence intervals.}
    \label{fig:Predicted_vs_True}
\end{figure}

To quantify the robustness of our results, we also test the NN predictions against a satellite galaxy sample from the Auriga suite of MW-mass zoom-in simulations. The Auriga project has better mass resolution than \eagle{} and uses a different galaxy formation model. We find that around 58\% of Auriga satellites have $t_{\rm infall}^{\rm true}$ within the 68\% CI, which is somewhat lower than we would expect from purely statistical considerations. This indicates that the NN, which was trained on \eagle{} data, performs somewhat worse when applied to the Auriga data.

This discrepancy is due to small systematic biases in the predicted infall time for the Auriga galaxies, which our NN predicts to have somewhat later infall times than actually measured in the Auriga simulations. This can be appreciated from  Figure \ref{fig:Predicted_vs_True}, which compares the NN inferred versus true infall times for a random subset of Auriga satellites. While most NN estimates agree with $t_{\rm infall}^{\rm true}$ within the shown 68\% CI, we find that on average it is more likely for the true infall time to be higher than the NN inferred maximum likelihood value. This can be seen in Figure \ref{fig:Predicted_vs_True} by noticing that more of the inferred maximum likelihood values, which are shown as filled blue circles, are below the one-to-one diagonal line shown in dashed grey.

The small discrepancy between the actual Auriga infall times and the ones predicted by the \eagle{}-trained NN are driven by differences in the halo mass profile predicted by the two projects. First, while MW-mass halos in \eagle{} loose a considerable fraction of their baryons due to strong supernovae feedback \citep{Halo_Energies_Schaller}, this effect is considerably reduced in the Auriga galaxy formation model \citep{Lovell2018}. Secondly, \eagle{} forms roughly a factor of two fewer stars in MW-mass halos compared to the \auriga{} predictions \citep{Main_Eagle_Schaye,Auriga}, which is due to the former model undershooting the stellar-to-halo-mass relation while the latter overshoots the same relation 
(see \citealt{Kelly2022} for a more detailed discussion of the differences between the \eagle{} and \auriga{} galaxy formation models).
Both these effects lead to \eagle{} halos having a shallower potential than their Auriga counterparts. This means that a satellite accreted at the same time in the two models ends up having different energies and, since the energy shows one of the strongest correlation to $t_{\rm infall}$, it explains why our NN shows a small systematic bias when applied to the Auriga data. One potential solution would be to use orbital action instead of energy, which has been shown to be a better orbital invariant under slow changes in the potential \citep[e.g.][although such   an approach will not mitigate all the differences between the two simulations, such as the different tidal disruption strengths due to the different stellar masses of the central galaxies, e.g. see \citealt{Richings2020DMvsEAGLE}]{Callingham2020}.

The gravitational potential differences between the \eagle{} and \auriga{} systems are largest close to the centre of haloes (which is where early accreted satellites have spent the most time) and decrease towards the outskirts. We find a similar trend when analysing the accuracy of the \auriga{} $t_{\rm infall}$ estimate. Farther satellites from the centre experience less systematic bias when inferring $t_{\rm infall}$. For example, the fraction of satellites that have $t_{\rm infall}^{\rm true}$ within the 68\% CI is 62.6\% and 66.6 \% for satellites with $r>0.5R_{200}$ and $r>0.75R_{200}$ respectively, which reconciles the \auriga{} and \eagle{} values. 

In summary, our comparison between the \eagle{} and \auriga{} samples shows that satellite infall time determinations can also be affected by systematic uncertainties that arise from mismatches in the host's gravitational potential and central galaxy stellar mass. This effect is most pronounced for satellites close to the centre, which corresponds on average to the earliest accreted satellites (e.g. see top-left panel in Figure \ref{fig:tinfall_vs_binned_Parameter_values}).

\section{Results for the MW satellites}
\label{chapter:results}

We now apply our two NNs to the observational data of the MW satellites. An important difference between determining infall times in simulations and doing so on real galaxies, is that we very precisely can determine the satellite properties in the former. However, in observations these properties have uncertainties. To obtain accurate confidence intervals, these uncertainties need to be taken into account. We do so by creating a sample of 1000 Monte Carlo (MC) realisations for the 6D phase-space coordinates of each observed satellite, as well as for the MW potential (see Section \ref{sec:obs_data} for details). All our predictions account for these uncertainties by averaging over the MC realisations and for some satellites we find that the observational uncertainties, in particular proper motion errors, can induce considerable further uncertainties on the infall time determination (for more details see Appendix \ref{appendix:mass_profile_table}).

\subsection{Satellites at first infall}
\label{sec:AON_results}

From the 47 MW satellites considered in this research, 44 have observed Galactocentric distances that are well within the MW halo radius, R$_{200}$, even after accounting for uncertainties in the MW mass profile. This means that most of the satellites in our sample can a priori be classified as already having been accreted by the MW, and thus subjected to environmental processes inside the MW halo.

When considering all the MC samples, there are six satellites that have MC realisations that lie outside of the MW halo. Theses are, in order of decreasing mean distance, Leo I, Leo II, Canes Venatici I, Leo V, Columba I, and Pisces II, and their mean distances from the Galactic Centre are 256, 234, 215, 194, 186 and $181~\rm{kpc}$, respectively (compare these with the MW halo radius, whose mean value is $218~\rm{kpc}$; \citealt{Cautun2020MW}). The predicted probabilities that these satellites have already been inside the MW radius are, in the same order as above, 82, 94, 98, 92, 88 and 97\%. Note that despite having a larger mean distance, Canes Venatici I has a larger accretion probability than Leo V and Columba I. The fact that all of the accretion probabilities are far above the 50\% mark, justifies the statement that all the satellites considered in this research have most likely already been accreted, even though there still is a significant probability (especially for Leo I and Columba I) that they might be at first infall.

This close to 100\% accretion score might raise questions on whether the observational data is correct, as for the \eagle{} sample a substantially lower score (about two thirds) was found. Most of the dichotomy can be explained by the fact that MW satellite surveys are still brightness limited and that previous Galactic surveys could only detect ultra-faint dwarfs, which account for most of the satellite population, if they were well within the halo of the MW. Recent studies predict that we have observed only half of the total MW satellites and that the yet-to-be-detected ultra-faint Galactic dwarfs are to be found predominantly in the outer regions of the Galactic halo \citep[e.g.][]{Newton2018TotalSats,Drlica-Wagner2020}. It is only with future generation telescopes that we should expect to find these faint and distant MW satellites, many of which will most likely not have been accreted yet.  

\begin{table}
\centering
\begin{tabular}{l l  l c l}
\hline
Name & \tinf &  \tquench & Ref. & \Dtquench  \\ 
 & [Gyrs] &  [Gyrs]  & & [Gyrs] \\[.05cm]
 \hline \hline  \\[-.25cm]
Sagittarius I & 8.5$^{+1.4, +3.4}_{-4.3, -7.5}$ & 3.4$^{+1.8}_{-0.3}$ & (1) & \textcolor{lightgray}{4.8$^{+1.4}_{-4.9}$}   \\
LMC & 0.7$^{+5.6, +9.2}_{-0.7, -0.7}$ &  -  & - & - \\
SMC & 0.7$^{+7.7, +10.1}_{-0.7, -0.7}$ &  -  & - & -  \\
Draco I & 9.4$^{+2.0, +2.9}_{-2.4, -6.7}$ & 9.1$^{+1.6}_{-3.3}$ & (1) & \textcolor{lightgray}{-1.4$^{+4.6}_{-2.4}$}  \\
Ursa Minor & 9.5$^{+1.7, +2.7}_{-2.5, -6.7}$ & 10.2$^{+1.5}_{-2.5}$ & (1) & \textcolor{lightgray}{-1.5$^{+2.7}_{-3.1}$}  \\
Sculptor & 8.7$^{+2.1, +3.6}_{-2.2, -7.0}$ & 10.6$^{+1.3}_{-3.5}$ & (1) & \textcolor{lightgray}{-3.0$^{+3.6}_{-3.0}$}  \\
Sextans & 1.0$^{+7.0, +9.2}_{-1.0, -1.0}$ &  -  & - & -  \\
Carina I & 8.5$^{+1.8, +3.5}_{-3.4, -7.8}$ & 2.2$^{+1.5}_{-0.0}$ & (1) & \textcolor{lightgray}{4.7$^{+3.2}_{-2.6}$}  \\
Fornax & 8.2$^{+2.2, +3.5}_{-2.7, -7.7}$ &  -  & - & -  \\
Leo II & 2.4$^{+5.4, +7.8}_{-1.4, -1.4}$ & 6.4$^{+0.8}_{-0.6}$ & (1) & \textcolor{lightgray}{1.3$^{+3.3}_{-4.4}$}  \\
Leo I & 1.7$^{+0.7, +6.3}_{-0.7, -1.7}$ & 1.7$^{+0.2}_{-0.1}$ & (1) & \textcolor{lightgray}{-0.5$^{+0.8}_{-0.7}$}  \\
Antlia II & 8.9$^{+1.5, +3.2}_{-2.5, -7.1}$ &  -  & - & -  \\
Aquarius II & 8.7$^{+2.2, +3.5}_{-3.2, -8.1}$ &  -  & - & -  \\
Bootes I & 8.7$^{+2.5, +3.6}_{-2.2, -6.6}$ & 12.6$^{+1.1}_{-1.0}$ & (1) & -3.1$^{+1.3}_{-3.7}$  \\
Bootes II & 0.7$^{+5.9, +9.4}_{-0.7, -0.7}$ &  -  & - & -  \\
Canes Venatici I & 7.7$^{+2.8, +3.6}_{-2.1, -6.3}$ & 8.3$^{+1.2}_{-2.0}$ & (1) & \textcolor{lightgray}{-1.6$^{+4.7}_{-2.2}$}  \\
Canes Venatici II & 8.8$^{+1.8, +3.2}_{-2.4, -6.6}$ & 12.7$^{+1.6}_{-1.6}$ & (1) & -3.3$^{+1.0}_{-3.6}$  \\
Carina II & 0.7$^{+0.8, +8.1}_{-0.7, -0.7}$ &  -  & - & -  \\
Carina III & 0.6$^{+0.7, +8.0}_{-0.6, -0.6}$ &  -  & - & -  \\
Columba I & 0.0$^{+6.3, +9.6}_{-0.0, -0.0}$ &  -  & - & -  \\
Coma Berenices I & 8.2$^{+2.9, +3.9}_{-3.6, -7.8}$ & 13.0$^{+1.3}_{-1.2}$ & (1) & -4.7$^{+2.8}_{-4.2}$  \\
Crater II & 8.9$^{+2.0, +3.2}_{-2.2, -6.6}$ &  -  & - & -  \\
Draco II & 7.4$^{+2.2, +3.9}_{-3.8, -7.0}$ &  -  & - & -  \\
Grus I & 0.4$^{+7.1, +9.6}_{-0.4, -0.4}$ &  -  & - & -  \\
Grus II & 8.4$^{+2.9, +3.5}_{-2.7, -6.6}$ &  -  & - & -  \\
Hercules & 1.0$^{+5.9, +8.8}_{-1.0, -1.0}$ & 11.8$^{+1.4}_{-1.3}$ & (1) & -11.3$^{+5.9}_{-1.9}$  \\
Horologium I & 8.4$^{+2.7, +3.8}_{-2.8, -7.8}$ & 11.5$^{+1.2}_{-1.1}$ & (2) & -3.1$^{+3.0}_{-3.3}$  \\
Horologium II & 8.9$^{+2.1, +2.9}_{-3.8, -8.5}$ &  -  & - & -  \\
Hydra II & 8.4$^{+2.7, +3.5}_{-3.6, -8.0}$ & 2.2$^{+0.3}_{-0.2}$ & (1) & \textcolor{lightgray}{6.8$^{+1.9}_{-3.9}$}  \\
Hydrus I & 0.7$^{+5.9, +9.2}_{-0.7, -0.7}$ &  -  & - & -  \\
Leo IV & 8.9$^{+1.5, +2.9}_{-2.9, -7.7}$ & 12.2$^{+1.4}_{-1.5}$ & (1) & -3.7$^{+1.9}_{-3.2}$  \\
Leo V & 7.5$^{+3.4, +3.8}_{-2.7, -7.3}$ &  -  & - & -  \\
Phoenix II & 0.6$^{+6.4, +9.6}_{-0.6, -0.6}$ &  12.5$^{+1.1}_{-1.1}$  & (2) & -11.5$^{+1.6}_{-0.7}$  \\
Pisces II & 0.1$^{+8.1, +10.5}_{-0.1, -0.1}$ &  -  & - & -  \\
Reticulum II & 8.1$^{+2.4, +3.5}_{-4.2, -6.6}$ & 12.3$^{+1.8}_{-1.8}$ & (2) & -5.1$^{+3.1}_{-3.5}$  \\
Reticulum III & 8.9$^{+2.1, +3.2}_{-3.2, -8.4}$ &  -  & - & -  \\
Sagittarius II & 0.8$^{+7.4, +9.8}_{-0.8, -0.8}$ &  -  & - & -  \\
Segue I & 8.2$^{+2.1, +3.6}_{-3.8, -7.3}$ &  -  & - & -  \\
Segue II & 7.7$^{+2.5, +3.6}_{-3.4, -5.7}$ &  -  & - & -  \\
Triangulum II & 8.2$^{+1.8, +3.6}_{-3.9, -7.7}$ & 12.9$^{+0.5}_{-0.8}$ & (2) & -4.9$^{+2.2}_{-3.6}$  \\
Tucana II & 0.7$^{+6.6, +9.6}_{-0.7, -0.7}$ & 12.8$^{+0.9}_{-0.8}$ & (2) & -12.5$^{+6.6}_{-1.3}$  \\
Tucana III & 3.6$^{+4.3, +7.4}_{-1.4, -2.0}$ &  -  & - & -  \\
Tucana IV & 8.2$^{+2.8, +3.6}_{-3.4, -6.8}$ &  -  & - & -  \\
Tucana V & 8.2$^{+2.7, +3.5}_{-3.9, -7.8}$ &  -  & - & -  \\
Ursa Major I & 9.4$^{+1.8, +2.9}_{-2.4, -7.0}$ & 11.2$^{+1.3}_{-1.2}$ & (1) & -3.1$^{+2.8}_{-2.1}$  \\
Ursa Major II & 8.2$^{+2.9, +3.6}_{-3.9, -7.8}$ &  -  & - & -  \\
Willman 1 & 8.9$^{+1.3, +2.5}_{-4.5, -7.0}$ &  -  & - & -  \\ \hline

\end{tabular}%
    \caption{
    The infall time, \tinf, and star formation quenching time, \tquench, of the MW satellites. The uncertainty ranges correspond to the 68 and the 95\% confidence limits; the latter is only given for \tinf{}. The quenching time denotes when those galaxies formed 90\% of their stars and are taken from: (1) \citet{Fillingham2019Quench} and (2) \citet{Sacchi_2021} (see 'Ref.' column). The last column gives the typical difference \Dtquench{} between the infall and quenching times of Galactic satellites. 
    The \Dtquench{} values for satellites with stellar mass, $M_\star\ge10^5\Msun$, are highlighted in light-grey since those are consistent with environmental quenching (see \reffig{fig:mytinf_vs_tquenchfill}).}
    \label{tab:tinf_vals}
\end{table}

\subsection{Accretion Time}
\label{sec:AT_results}

Based on the assumption that indeed all the satellites are accreted, we proceed to determine their infall time distribution. The resulting PDFs are show in Appendix \ref{appendix:mass_profile_table} and the most likely estimates and the 68 and 95\% CI are summarised in Table \ref{tab:tinf_vals}. We advise a note of caution when interpreting these values: most of the inferred \tinf{} likelihoods are highly asymmetrical and many show two or more peaks, such as Sextans and Leo II (see Figure \ref{fig:tinf_results_1}). Moreover, a considerable fraction of satellites, such as Carina II, Hydrus I and Hercules, have \tinf{} likelihoods that show a high and narrow peak at recent lookback times and a long tail (and even a second peak) towards earlier infall times. Keeping the above in mind, one should always have a look at the actual inferred likelihoods when interpreting the size of the CI or when comparing our infall times with other measurements.

\section{Discussion}
\label{chapter:Discussion}

We now proceed to interpret the infall times we inferred for the Galactic satellites. We will focus on a few questions, ranging from how typical are the accretion times of MW satellites when compared with theoretical predictions for MW-sized systems to what are the implications for determining the star-formation quenching timescale in the Galactic halo.  

\subsection{Comparing Galactic infall times to theoretical predictions}
\label{sec:MWvsEagle}
\begin{figure}
    \centering
    \includegraphics[width = 1.0\linewidth]{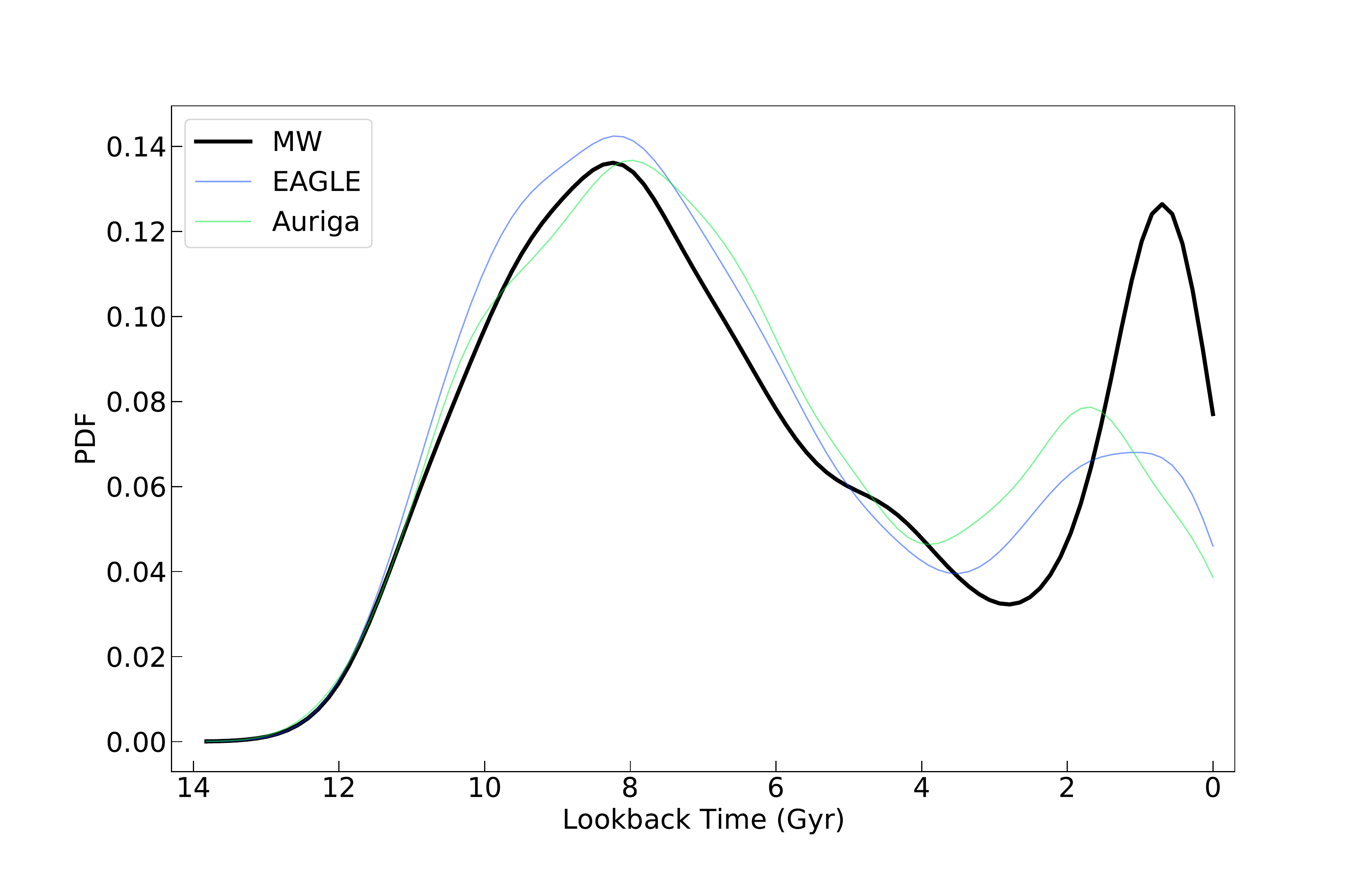}
    \vskip -.2cm
    \caption{Combined infall time likelihood for all the MW satellites (black line). This is compared with the \tinf{} distribution for the population of satellites of MW analogues in \eagle{} (blue) and Auriga (green). For the simulations, we show infall time for $z=0$ satellites found within $1.4R_{200}$ from the host, which is equivalent to within a distance of $300~\rm{kpc}$ for a system with MW's total mass.}
    \label{fig:MW_vs_EAGLE}
\end{figure}

The first question raised by our results is how do the accretion times of Galactic satellites compare to theoretical expectations. For the former, we combine the \tinf{} likelihoods from each of the 47 MW dwarf studied here to obtain the \tinf{} likelihood for the population of observed MW satellites, which is shown in Figure \ref{fig:MW_vs_EAGLE}. We remind the reader that the observed MW dwarfs are an incomplete radially-biased sample since they are found in brightness limited surveys. In Figure \ref{fig:tinfall_vs_binned_Parameter_values} we show that the infall time shows a weak correlation with the present-day radial distance, with the median \tinf{} increasing by ${\sim}2~\rm{Gyrs}$ between dwarfs at $0.2R_{200}$ and those at $R_{200}$. This suggests that the observed dwarfs might be biased towards earlier infall times compared to the full population of Galactic satellites.

Figure \ref{fig:MW_vs_EAGLE} shows that most of the currently observed Galactic satellites were accreted at early times, with a broad peak around $8~\rm{Gyrs}$ ago (i.e. redshift $z=1$), with the oldest satellite having orbited in the MW halo for around $11~\rm{Gyrs}$. The MW satellite accretion rate decreases towards present time except for a high and narrow peak at $1.5~\rm{Gyrs}$ ago. This second peak is due to the accretion of the LMC, which is currently at its first infall onto the MW \citep[e.g.][]{Besla2007,Cautun2019LMC}. Due to its relatively large mass, the LMC also hosts its own satellite galaxies \citep[e.g.][]{Jethwa2016,Patel2020LMCSMC}. It is therefore a reassuring sight that our results shows a peak in infall probability around the 1.5 Gyr mark, which is when the LMC would have entered the MW halo (see Figure \ref{fig:tinf_results_1}), together with its satellites. We discuss the LMC satellites in more details in Section \ref{sec:MCdisc}.  

To compare with theoretical expectations, we calculate the infall time distribution for the \eagle{} and Auriga satellites of MW-analogues, which we limit to satellites within 1.4 R$_{200}$ which corresponds to objects within $300~\rm{kpc}$ for the fiducial MW halo mass. Figure \ref{fig:MW_vs_EAGLE} shows the resulting distributions for the two galaxy formation models. Overall, we find good agreement between the \tinf{} PDF measured in \eagle{} and Auriga, although some small discrepancies are present that are likely due to the differences in host potential that we discussed in Section \ref{sec:performance_AT}. 
The \tinf{} PDF is bimodal, with a second peak at around $2~\rm{Gyrs}$ lookback time. The second peak is due to the fact that many satellites accreted around $4~\rm{Gyrs}$ ago are presently found at distances larger than the 1.4 R$_{200}$ threshold value used in Figure \ref{fig:MW_vs_EAGLE}. Increasing this distance to 2 R$_{200}$ nearly removes the second peak (see Figure \ref{fig:VMax_vs_tinfall}) by mostly adding satellites with infall times, $\tinf\in[2,6]~\rm{Gyrs}$ (see \citealt{Simpson2018} for a more detailed analysis).

At early times, we find good agreement in the \tinf{} likelihood between the MW and theoretical prediction indicating that the early accretion of dwarfs onto the MW is typical of $\Lambda$CDM predictions. It is only around $3~\rm{Gyrs}$ lookback time that the MW curve starts to deviate strongly, first below the EAGLE and Auriga predictions and then increasing to a sharp and high peak, which we interpreted as the accretion of the LMC and its satellites. 

The MW is predicted to have had a few more massive satellite accretions besides the LMC \citep[e.g.][]{Kruijssen2020,Callingham2022}, and each such massive accretion is expected to bring at the same time a surplus of satellites \citep{dsouza2021massive}. Two such events are the Gaia-Enceladus-Sausage (\citealt{Belokurov2018Sausage,Helmi2018Sausage}) and Kraken \citep{Kruijssen2019,Kruijssen2020}, which, while uncertain, are believed to have had stellar masses nearly as high as the LMC one and to have been accreted $8-11~\rm{Gyrs}$ ago. However, the MW \tinf{} PDF does not show one or more significant peaks at early times except the main and very broad peak at $8~\rm{Gyrs}$ ago that is nearly the same as when averaging over all MW-analogues in the EAGLE and Auriga samples. The broadness of the peak rather suggests a more steady accretion of multiple small satellites, a theory for which hints were found by \citet{Wang2021MultipleSausage}. The reader should recall however that, as discussed in section \ref{sec:performance_AT}, early \tinf{} determinations are also the most uncertain and that could potentially dampen any early massive accretion peaks.

\subsection{Comparing with previous infall time determinations}
\label{sec:me_vs_fill}

\begin{figure*}
    \centering
    \mbox{\hskip -1.8cm \includegraphics[width = 1.2\linewidth, height = 0.5\linewidth]{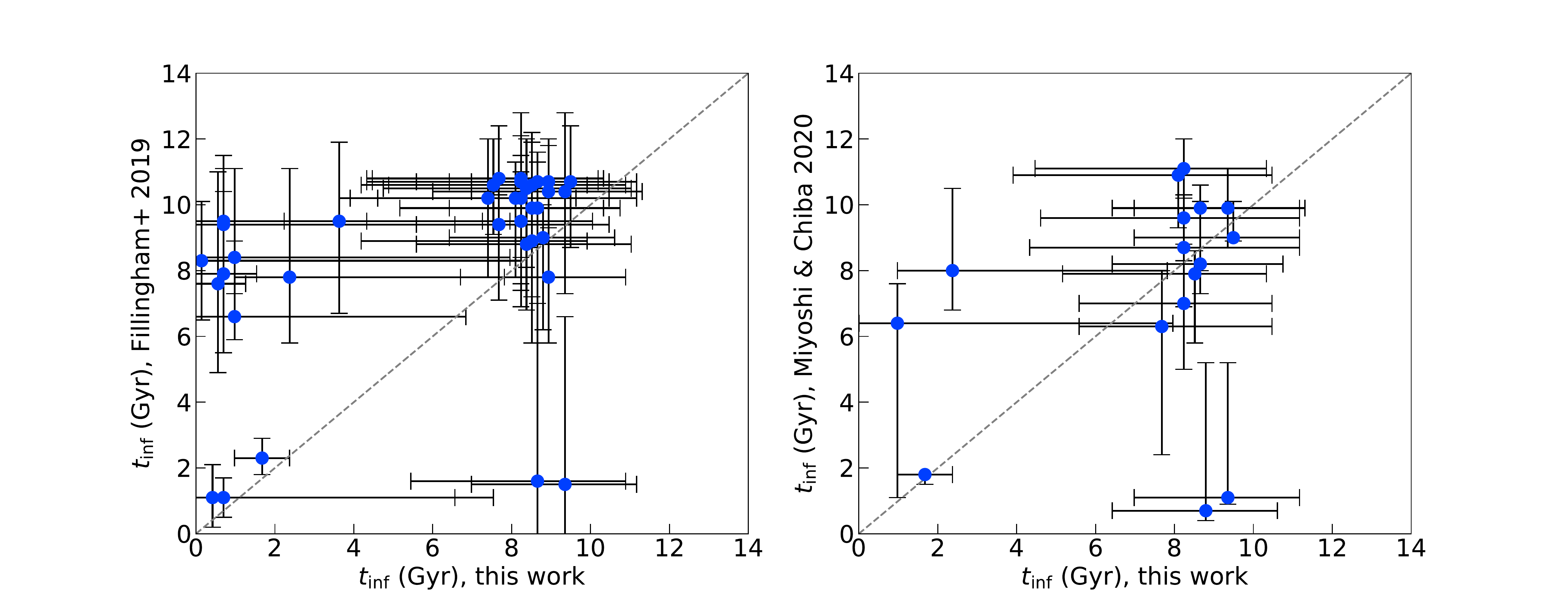}}
    \caption{Left panel: comparison between the infall times found in this work and the infall times found in \citet{Fillingham2019Quench}. The diagonal dotted line is the one-to-one line. Right panel: same as left panel, but now with the infall times as found in \citet{Miyoshi2020Infall1st} instead of \citet{Fillingham2019Quench}.}
    \label{fig:mytinf_vs_tinfthem}
\end{figure*}

We now compare the infall times found in this work to earlier studies. As mentioned in the introduction, \citet{Fillingham2019Quench} determined the infall times for a sample of the MW satellites by comparing orbital properties, in particular the satellite energy, to satellites in simulations. They found infall times for 37 satellites, all of which are also considered in this work. The comparison with the \citeauthor{Fillingham2019Quench} results is shown in left-hand panel of Figure \ref{fig:mytinf_vs_tinfthem}.

Before discussing the results, one should realise that there are multiple differences in both data, simulations, and methodology between our work and the \citeauthor{Fillingham2019Quench} one. Most importantly, our work uses the updated satellite proper motions from \citet{Mcconnachie2020propermotions}, whereas \citeauthor{Fillingham2019Quench} have relied on more uncertain proper motions from \citet{Fritz2018Proper} based on {\it Gaia} DR2. Secondly, \citeauthor{Fillingham2019Quench} employ dark matter only simulations, while we use hydrodynamic ones that include models for most of the processes thought to be important for galaxy formation. In particular, such processes can lead to change in the gravitational potential of the host halo, which affect both the infall time and orbits of satellites, as well as the tidal disruption of satellites, which is enhanced in simulations containing stellar discs \citep[e.g.][]{Sawala2017,Richings2020DMvsEAGLE,Green2022}.   

Keeping the above in mind, a couple of conclusions can be inferred when comparing our results with the \citeauthor{Fillingham2019Quench} ones.
Firstly, while the majority of the infall times agree within the 68\% CI (23 out of 37, which represents 62\% of the common sample), a sizeable minority (clustered in the top left corner of the figure) is in rather stark disagreement. It can be said that from the current LMC accretion onto the MW (see section \ref{sec:MCdisc}), one would expect a relatively large group of satellites at recent infall times. The disagreeing cluster would agree with this group in our results, while only a hand full of the satellites in \citeauthor{Fillingham2019Quench} show such recent infall times. The lack of such recent infall times in \citeauthor{Fillingham2019Quench} might be explained by the fact that the twelve high resolution simulations used by the authors did not contain any LMC-size satellite. This can be determined from Figure 2 in \citet{PhatELVIS2019}, which shows that no subhaloes have V$_{\rm peak} > 40~\rm{km/s}$. Noting that the LMC has a stellar mass well above 10$^{9}$ M$_{\odot}$ \citep{Mcconnachie2012SatCoords}, its V$_{\rm peak}$ value would be closer to $70-90~\rm{km/s}$ \citep{Fattahi2018}. In addition, the lack of stellar disc disruption in the dark matter only simulations employed by \citeauthor{Fillingham2019Quench} leads to subhaloes in the inner regions of their host surviving for longer and thus will also lead to more satellites with early infall times. 

Secondly, almost all the data points in the plot lie above the one-to-one line, even when not considering the top-left cluster. This means that the infall times found by the NN are almost always slightly more recent than those from \citeauthor{Fillingham2019Quench}. The most likely explanation for this observation has already been discussed in section \ref{sec:performance_AT}, namely that the NN predictions tend towards the mean value when the predictions are highly uncertain. This will inevitably bias the maximum likelihood estimates (but not the confidence intervals) slightly towards more recent infall times, i.e. above the one-to-one line in the figure.

Another observation is that there is a dearth of satellites with lookback infall times between roughly 2 and 8 Gyrs in both samples. While it could be argued from Figure \ref{fig:MW_vs_EAGLE} that the theoretical predictions for the probability of infall in this time range is somewhat lower than at earlier times, nonetheless this range should contain a significant fraction of satellites. One possible explanation is the existence of an observational bias: many galaxies falling in during this time frame are currently around their first or second apocentre \citep{dsouza2021massive}, making them harder to observe and thus less likely to be present in the MW sample of satellites  which is magnitude limited. 

In the right-hand side panel of Figure \ref{fig:mytinf_vs_tinfthem} we compare our results with those of \citet{Miyoshi2020Infall1st}, which have used a very different method for determining infall times. That study has employed backwards integration of the satellites in a time dependent MW potential. As the numbers are not present in the paper, a request for the infall time data from \citeauthor{Miyoshi2020Infall1st} was sent out to the paper's authors, who kindly provided us their data. 

On first glance, the agreement is a lot better here: the outlying top left cluster found in the comparison with the \citeauthor{Fillingham2019Quench} results is mostly gone, and the infall times of only 5 out of 16 (31\%) satellites do not agree within the 68\% CI. Some notes of caution are necessary however. The \citeauthor{Miyoshi2020Infall1st} sample is smaller and includes only 16 of the 37 satellites from the \citeauthor{Fillingham2019Quench} study, with the other objects being excluded due to either being potential LMC satellites or due to having large proper motion uncertainties. Out of the 14 satellites showing a large disagreement between our and the \citeauthor{Fillingham2019Quench} results, only 3 are present in the \citeauthor{Miyoshi2020Infall1st} sample, and 2 out of these 3 also disagree with our results within the 68\% CI.

Arguably the most important take away from the comparison of our \tinf{} results with those of previous studies is that determining infall times for satellites remains a far from easy task. Typical confidence intervals for all studies are large at almost half the age of the Universe and the agreement between contemporary studies is poor.

\subsection{The Magellanic Satellites}
\label{sec:MCdisc}

As discussed before, some of the satellites considered in this research are thought to have been LMC satellites when they were accreted onto the MW \citep[e.g.][]{Jethwa2016,Kali2018LMC,Patel2020LMCSMC,Ethan2021LMCQuenching}. While by no means definitive, the infall time distribution can be a valuable indicator to determine the likelihood of the association between satellites and the LMC. According to \citet{Patel2020LMCSMC}, long-term LMC satellites are the SMC, Carina II, Carina III, Horologium I and Hydrus I, while Phoenix II and Reticulum II are recently accreted by the LMC. The first four were already proposed in an earlier work \citep{Kali2018LMC}, which also mentions Reticulum II, Draco II, Tucana II, Hydra II and Grus I as less likely companions. These satellites are indicated in appendix \ref{appendix:mass_profile_table}, where we give the individual \tinf{} PDF of each satellite, by having their name in red for the most probable and in orange for the less probable ones.

Three of the most likely LMC companions, Carina II, Carina III and Hydrus I, have infall time distributions with a similar shape to that of the LMC and with a considerable likelihood to have been accreted around $1.5~\rm{Gyrs}$ ago. The other two likely LMC satellites, SMC and Horologium I, also have a likelihood peak at $\tinf=1.5~\rm{Gyrs}$, however the NN predicts that they are more likely to have have been accreted earlier, around $\tinf\simeq 8~\rm{Gyrs}$. We suspect this discrepancy is due to not including the LMC potential when modelling the MW, with the LMC thought to have been rather massive at infall \citep{Penarrubia2016,Cautun2019LMC} and even today having a considerable total mass \citep{Garavito-Camargo2019,Erkal} that can have a large impact on the motion of dwarfs close to the LMC such as the SMC and Horologium I \citep[e.g. see][and also \citealt{Erkal} for example of Galactic streams]{Patel2020LMCSMC}. We obtain a similar picture when studying the less likely LMC satellites,  with Phoenix II, Tucana II, and Grus I having \tinf{} PDFs very similar to that of the LMC, while Reticulum II, Draco II, and Hydra II show more uncertain \tinf{} determinations with a large probability of early infall too.

It is reassuring to find that our NN predicts that many probable LMC satellites have similar infall times as the LMC. This is a non-trivial results since our MW model does not include information about the potential of the LMC or the distance of the satellites from the LMC. This result suggests that, as expected, for most satellites the MW potential is the dominant one and that the LMC contribution can be neglected to a first approximation. However, this is not the case for dwarfs close to the LMC, such as the SMC and Horologium I \citep{Garavito-Camargo2021}. 

A potential improvement to the present work would be to add the LMC potential, or, more generally, the potential of massive satellites.  In doing so, we would further solidify the infall time distributions of especially the satellites that currently are close to the MC's, allowing for more robust claims on LMC and SMC association. One simple approach to do so within our current NN framework would be to constrain our sample to host galaxies that have an LMC analogue. However, LMC mass satellites are quite rare for MW like hosts \citep[e.g.][]{Shao2018b}. Besides the resulting drop in sample size, another difficulty would be to define when a satellite can be considered an LMC analogue. Mass and distance from the host are two important criteria, but potentially many more such as the orbit and the number of pericentre passages. Due to their limited volume, current hydrodynamical simulations such as \eagle{} do not allow for sufficient MW and LMC analogue pairs to extend the NN framework to also include the LMC potential.   

\subsection{Quenching timescale}
\label{sec:Quenching}
\begin{figure*}
    \centering
    \includegraphics[width = 1\linewidth, height = 0.5\linewidth]{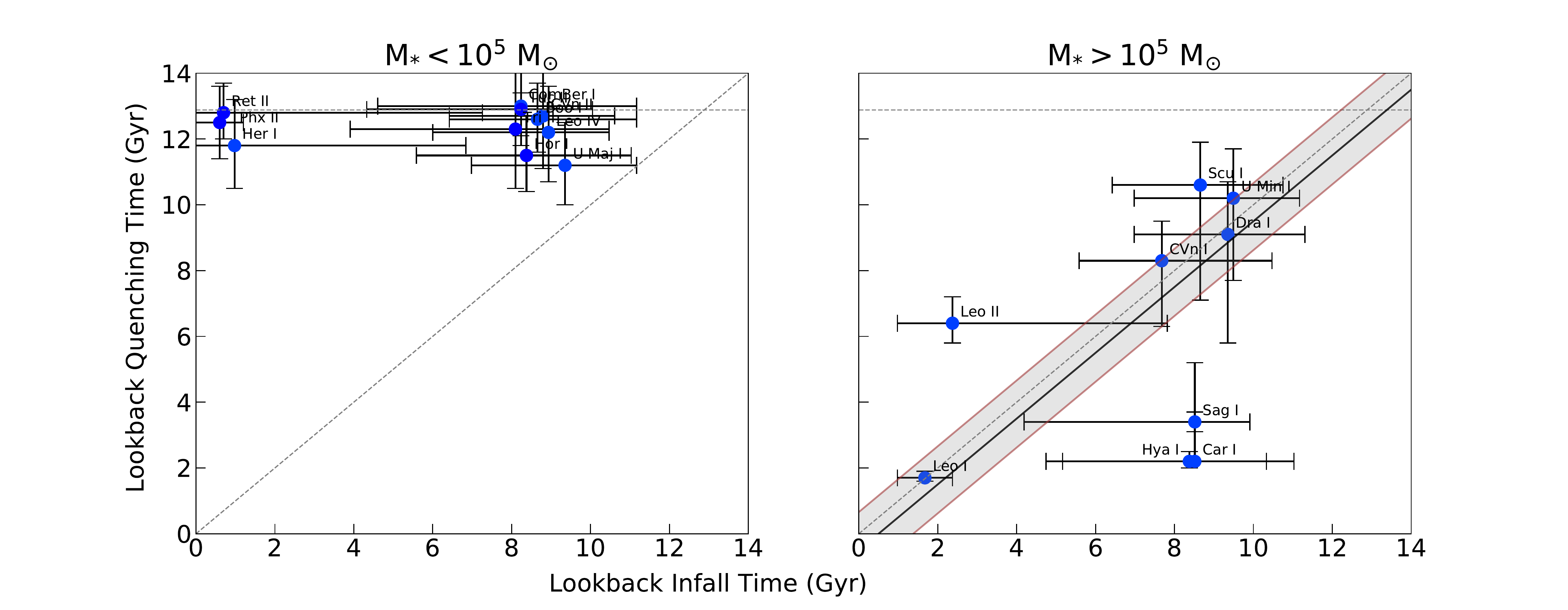}
    \caption{
    Quenching time versus infall time for 20 MW satellites. The infall times are as determined in this research, the quenching times are from \citet{Fillingham2019Quench} and \citet{Sacchi_2021}. The Galactic dwarfs are split into two according to their stellar mass, with M$_\star < 10^{5}$ M$_{\odot}$ (i.e. UFDs) shown in the left panel and M$_\star > 10^{5}$ M$_{\odot}$ (i.e. classical dwarfs) shown in the right panel. The end of reionisation at $z\simeq6$ (a lookback time of $12.9~\rm{Gyrs}$) is indicated by a horizontal dotted line. The diagonal dotted line is the one to one line. In the right panel, the grey shaded region is the 68\% confidence interval on the average quenching timescale ($t_{inf} - t_{q; 90}$), with the black solid line the maximum likelihood estimate.}
    \label{fig:mytinf_vs_tquenchfill}
\end{figure*}

Most environmental star-formation quenching studies follow a statistical approach that connect the fraction of quiescent galaxies with the mean accretion history of satellites to obtain an average quenching time \citep[e.g.][]{Wetzel2013,Wetzel2014,Slater2014,Fillingham2015}. However, the wealth of Galactic data, where we can determine star-formation histories and infall times for individual satellites, allows for the complementary approach of studying the correlation between quenching and accretion for each satellite \citep[e.g.][]{Fillingham2019Quench}. Here we follow this latter approach and analyse the relation between our inferred infall times and the quenching times for 20 Galactic satellites as provided in  \citet{Fillingham2019Quench} and \citet{Sacchi_2021}. This dependence is shown in Figure \ref{fig:mytinf_vs_tquenchfill}.

For each Galactic satellite with available data we calculate the quenching timescales, $\Dtquench{} =  \tinf{} - \tquench{}$, where \tquench{} is the time when a galaxy formed 90\% of its present-day stars. To determine the \Dtquench{} uncertainties, we use an MC approach. For \tinf{} we take  the MC samples that are outputted by our analysis. For \tquench{} we only have access to the MLE and the 68\% CI (see third column in Table \ref{tab:tinf_vals}) and, since the CI is not symmetric around the MLE, to generate the MC samples we approximate the \tquench{} likelihood as the composite of two Gaussians. One Gaussian describes the distribution of \tquench{} values below the MLE, with the mean and dispersion of this distribution being given by the MLE and the absolute difference between the lower end of the 68\% CI and the MLE. The second Gaussian models the likelihood of \tquench{} values above the MLE, with mean equal to the MLE and dispersion given by the difference between the upper end of the 68\% CI and the MLE. The resulting quenching timescales and their 68\% CI are given in the last column of Table \ref{tab:tinf_vals}.
 
As discussed at length in section \ref{sec:introduction}, we expect a dichotomy in the quenching processes between massive dwarfs, with stellar mass $M_\star>10^5\Msun$, and lower mass ones, with $M_\star<10^5\Msun$. Motivated by this theoretical expectation, we separate Figure \ref{fig:mytinf_vs_tquenchfill} into two panels that show the \tinf{} versus quenching time relation for low $M_\star$ (left panel) and high $M_\star$ (right panel) Galactic dwarfs. We find that the ultrafaint dwarfs show no clear correlation between infall time and quenching time. This is to be expected from a scenario in which quenching due to reionisation is dominant \citep{Okamoto2008Reionisation, Bose2018Reionisation} since reionisation, which happened at $z>6$ \citep{Planck2016_reionization}, took place before all present-day surviving MW satellites were satellites of another halo \citep{Wetzel2015}. The left panel of Figure \ref{fig:mytinf_vs_tquenchfill} shows that within uncertainties all low stellar mass dwarfs are consistent with having been quenched at roughly the same time about $12 - 13~\rm{Gyrs}$ ago. None of the ultrafaint dwarfs have a quenching time more recent than their infall time, even when accounting for the 68\% CI (see the \Dtquench{} values in Table \ref{tab:tinf_vals}).

The right panel in Figure \ref{fig:mytinf_vs_tquenchfill} shows that classical dwarfs show a strong correlation between infall and quenching times, with all systems being compatible with $\tinf{} > \tquench{}$ within the 68\% CI\footnote{One could argue that Leo II is an exception, but when looking at the infall distribution for Leo II (see figure \ref{fig:tinf_results_1}) we find that its exact infall time is highly uncertain, with practically equal probability for 2 Gyr or 8 Gyr ago.}. We find that for most dwarfs quenching takes place basically at accretion, and for three of them, Sagittarius I, Hydra II, and Carina I, at about $5~\rm{Gyrs}$ after infall, although the \tinf{} uncertainties are rather large and we cannot exclude a short quenching timescale also for these last three dwarfs. Differences in the quenching times of individual satellites are expected for ram-pressure stripping, which has been proposed as the dominant process \citep{Fillingham2015,Fillingham2016}, since the removal of cold gas depends on the balance between the ram-pressure and the restoring forces, which depend on the orbit and total mass of a satellite \citep{Simpson2018}.

When averaging over the results for all the massive dwarfs in our sample, the quenching timescale is found to be $0.5^{+0.9}_{-1.2}$ Gyrs (68\% CI). This is in good agreement with previous results that find $\Dtquench \sim 0-2~\rm{Gyrs}$ \citep{Slater2014,Fillingham2015,Fillingham2019Quench}. This relatively quick quenching further strengthens the theory of environmental quenching, and means that, at least for dwarf satellites with mass $M_\star>10^5\Msun$, the quenching time is a good proxy for the infall time.

\section{Conclusions}
\label{chapter:conclusions}

We have developed a neural network (NN) for inferring the infall time of satellites of MW-mass systems that we have trained using MW-analogues in the \eagle{} project, which is a cosmological hydrodynamical simulation that reproduces many key properties of galaxies in the observed Universe. The NN takes as input the phase-space coordinates of satellites and the specific energy, with the latter showing the largest correlation with the infall time \citep[see also][]{Rocha2012Infall2nd}. We further scale these properties by quantities proportional to the host halo mass since satellite properties have been shown to be universal when scaled appropriately \citep[e.g.][]{Callingham2019Scaling}. The NN has been designed to predict the \tinf{} likelihood for each individual satellite without making assumptions on the shape of this function, which we achieve by predicting the likelihood in many equally spaced time intervals that span the age of the Universe.

We have tested the NN prediction using a test subset from the \eagle{} project and another independent set from the \auriga{} suite of simulations to find that our NN predicts realistic confidence intervals. In the latter case, we found that our uncertainties were slightly too low, which we traced back to a small systematic bias in the inferred \tinf{} values for the Auriga satellite galaxies. This is due to differences in the mass profile of the central host, whose potential is shallower in \eagle{} than in \auriga{}. 
This means that on top of the statistical errors that we have quoted, our results are affected also by small systematic uncertainties that are most pronounced for satellites close to the centre. To fully quantify these systematic uncertainties we would need to analyse a larger number of galaxy formation models than the two employed here.

We have applied the NN to 47 MW dwarf galaxies with both 3D positions and velocities that are found within a distance of $300~\rm{kps}$ from the Galactic Centre. Since this distance is larger than the Galatic $R_{200}\simeq220~\rm{kpc}$, which we take as the extent of the Galactic halo, we have developed a second NN that predicts if a dwarf found at a distance larger than $R_{200}$ is at first infall or is a backsplash galaxy, i.e. a satellite that already had a pericentre passage closer than $R_{200}$ and that is on an extended orbit which takes it outside its host halo. This second NN achieves a better than 85\% accuracy of distinguishing between first infall and backsplash galaxies.

The main conclusions of our study are as follows:

\begin{itemize}
    \item Our NN predicts infall times with an average 68\% confidence interval of size 4.4 Gyrs. This uncertainty can be considerably lower for recently accreted satellites and somewhat larger for early accreted ones. 
    \item All the MW satellites considered in this work are very likely to have entered the Galactic halo and thus experienced environmental effects, even the ones currently found at distances larger than $R_{200}$. The lowest backsplash probability is 82\% for Leo I, and it is higher than 90\% for the other five dwarfs that potentially could lie outside the $R_{200}$ radius.
    \item The infall time distribution of MW satellites follows the average predictions of the \eagle{} and \auriga{} models with one difference. The MW shows a second narrow peak in the \tinf{} likelihood at a $1.5~\rm{Gyrs}$ lookback time that we associate to the accretion of the LMC and its satellites.
    \item For many of the dwarfs that have been proposed as LMC satellites we find \tinf{} likelihoods very similar to that of the LMC even though our Galactic model does not include a massive LMC component. This find illustrates the robustness of our results and that neglecting the LMC potential is a reasonable first approximation. However, for the SMC and Horologium I, we find considerably earlier accretion times than the LMC, indicating that for dwarfs close to the LMC we cannot neglect the potential of this massive satellite.
    \item We have compared our \tinf{} determination with the backward orbital integration of \citet{Miyoshi2020Infall1st} to find reasonable agreement. The comparison with the \citet{Fillingham2019Quench} infall times showed a mixed picture, with good agreement for a significant fraction of satellites, but large discrepancies with the presumed LMC satellites that \citeauthor{Fillingham2019Quench} predicts to have been accreted considerably earlier than the LMC.
    \item We have also studied the correlation between infall time and star-formation quenching times. These are unrelated for dwarfs with stellar masses $M_\star<10^5\Msun$, indicating that reionisation was the dominant quenching process for these low mass galaxies. For higher stellar masses, we find a considerable correlation between accretion and quenching, with star-formation ending on average very shortly, $0.5^{+0.9}_{-1.2}$ Gyrs (68\% CI), after a satellite crosses the $R_{200}$ radius. 
\end{itemize}
    
Our work has shown that NN can be used to solve a challenging cosmological problem: how to infer the accretion time of satellites from present-day observables? The use of NN has the advantage of going beyond simplified models of satellite motions, such as those employed in backwards orbit integration, and offers a natural way of connecting satellite orbits in observations with their counterparts in cosmological simulations. To further advance this work, one would need to add the gravitational potential of massive satellites, such as the LMC, and possibly use the orbital actions instead of the energy and angular momentum as input NN parameters. Orbital actions are better conversed than the energy (\citealt{Callingham2020} shows this for actual MW-mass simulations) and potentially would be more strongly correlated to the infall time, especially for early accretion events. Having a larger training sample would also be helpfull in increasing the number of parameters used when training the NN. Currently, we use only orbital information, but additional information could be satellite colours or the quenching time, for which we find a strong correlation with infall time. However, going beyond orbital parameters should be done with care and only once we have a better understanding of galaxy formation physics and how it relates to galaxy orbits.

\section*{Acknowledgements}
 
We thank the anonymous referee for their constructive comments, helping us to improve the paper.
We thank Matthieu Schaller for giving us the best-fitting parameters for the total mass profile of \eagle{} haloes. We also thank Robert J. J. Grand for very helpful comments and for providing access to the \auriga{} suite of simulations.
MC acknowledges support by the EU Horizon 2020 research and innovation programme under a Marie Sk{\l}odowska-Curie grant agreement 794474 (DancingGalaxies).
This work used the DiRAC@Durham facility managed by the Institute for Computational Cosmology on behalf of the STFC DiRAC HPC Facility (www.dirac.ac.uk). The equipment was funded by BEIS capital funding via STFC capital grants ST/K00042X/1, ST/P002293/1, ST/R002371/1 and ST/S002502/1, Durham University and STFC operations grant ST/R000832/1. DiRAC is part of the National e-Infrastructure.

\section*{Data availability}
This paper made use of the publicly available data from the \eagle{} database: \url{http://icc.dur.ac.uk/Eagle/database.php}. We tested our method using satellite infall times from the \auriga{} suite of simulations that can be made available upon reasonable request to the authors. 

\bibliographystyle{mnras}
\bibliography{references}

\appendix

\section{NN Architecture}
\label{appendix:NN_architecture}

The two MLP networks used for this research were built using the \texttt{scikit-learn} python package \citep{scikit-learn} and used the cross-entropy as the cost function being minimised during the training stage. The cross-entropy measures the difference between two distributions and is a widely used loss function for classification models. Each NN has three hidden layers with 100, 80 and 60 neurons, respectively, that has been found by testing various NN architectures and choosing the one with the minimum number of layers and neurons such that, when increasing this further, does not result in an improvement in the loss function of the evaluation sample. We have used the early stopping option, which means that the training was stopped once the validation score did no longer improve. We have made a 60\%-20\%-20\% split between the training-, validation- and test-samples. We have used the \texttt{adam} optimiser and tested different learning rates and found that the optimal value was 0.001; values close to this did not have a large effect on the cost function, although very large or low values did lead to worse predictions. 

Using the architecture described above, we built two MLPs networks: i) one for determining if a satellite is at first infall, and ii) for predicting the infall time likelihood. Choosing the same architecture for both MLPs is justified as it was found that increasing the numbers of hidden layers and neurons per layer did not improve the prediction for either of the two models, while a considerably simpler network did lead to worse predictions. The first MLP was built to determine whether or not a dwarf galaxy outside the host's virial radius is at first infall or actually a backsplash satellite. The second MLP was used to determine, assuming that a satellite has fallen in, at what time it fell in. 

\section{Individual Infall Time Distributions}
\label{appendix:mass_profile_table}

\begin{figure*}
    \hskip -3.7cm
    \centering
    \includegraphics[width = 1.2\linewidth, height = 1.2\linewidth]{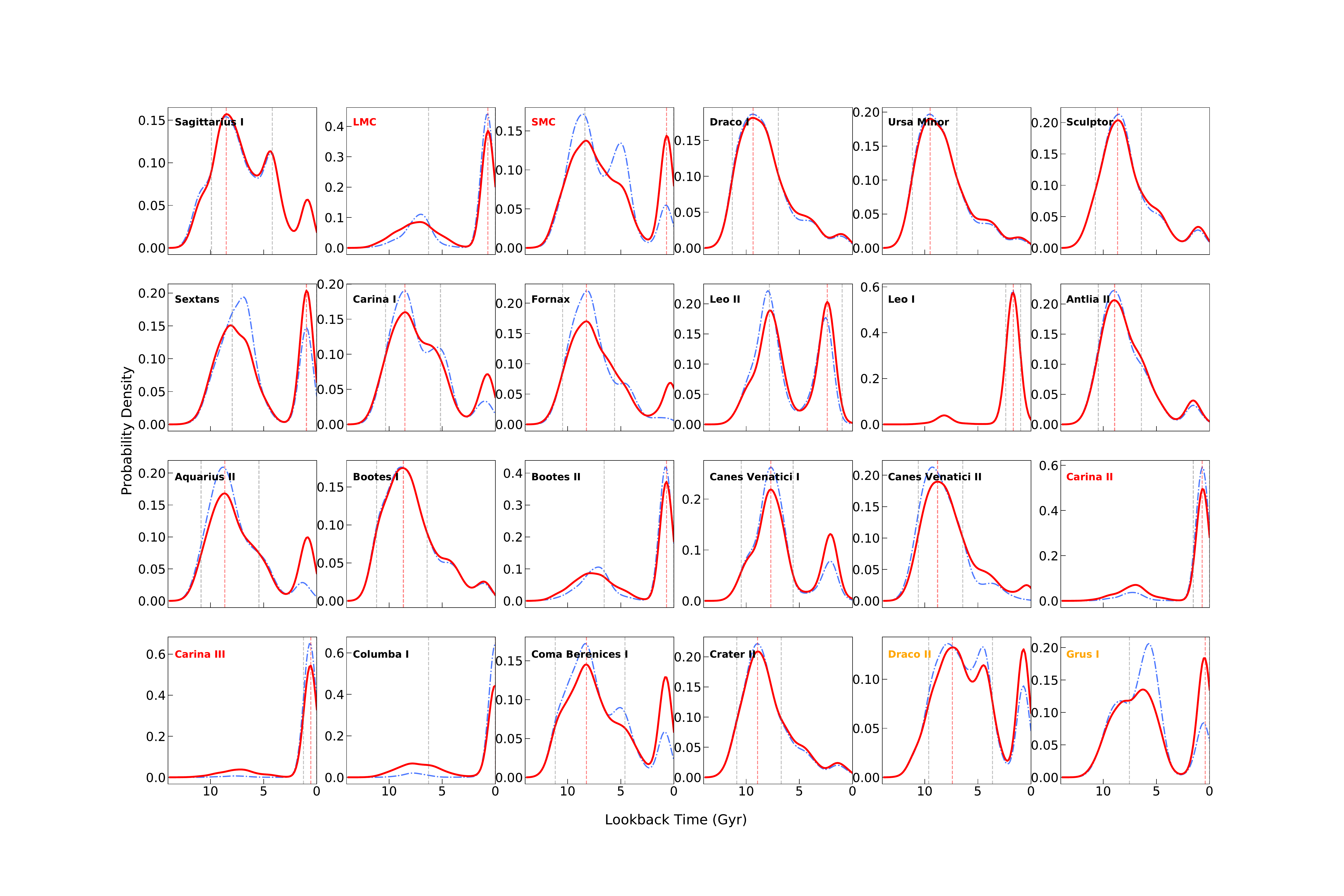}
    \hskip -3.5cm
    \caption{ The the infall time likelihood for 24 MW satellites. The fiducial prediction is shown by the red lines and show the results when accounting for errors in the measured properties of Galactic satellites and in the MW mass profile. The vertical red dashed line shows the most likely infall time and the two vertical grey dashed lines show the 16 to 84 percentiles. Satellites that are considered to be long-term LMC satellites according to \citet{Patel2020LMCSMC} and \citet{Kali2018LMC} have their name in red, while those with their names in orange are considered to be recently accreted/less likely LMC satellites (see section \ref{sec:MCdisc}).
    To illustrate the effect of measurement uncertainties, we also show using the blue dashed-dotted line the infall time likelihood but now assuming the ML position and velocity of satellites and the ML MW mass profile. }
    \label{fig:tinf_results_1}
\end{figure*}

\begin{figure*}
    \centering
    \hskip - 3.7cm
    \includegraphics[width = 1.2\linewidth, height = 1.2\linewidth]{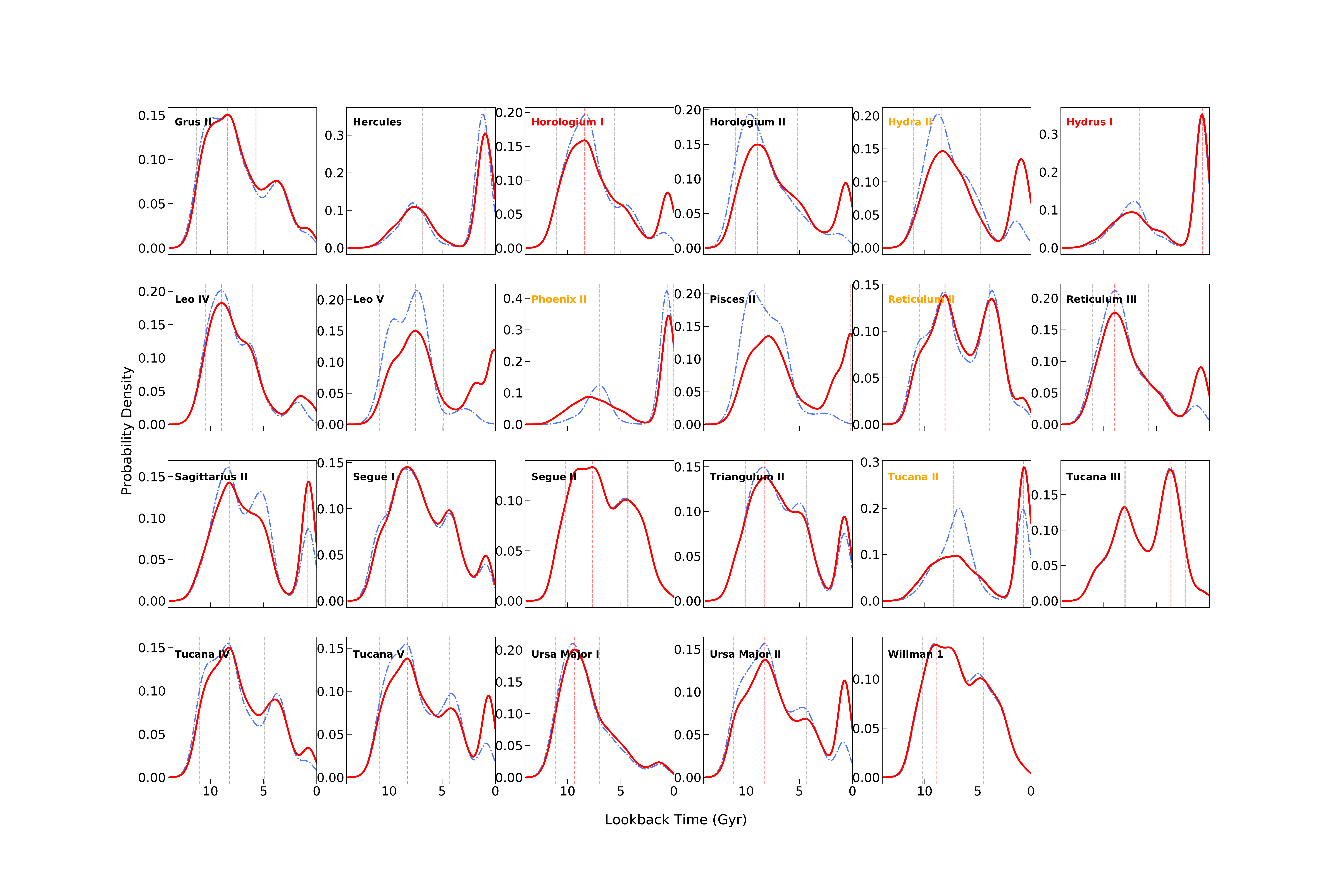}
    \hskip -3.5cm
    \caption{Same as figure \ref{fig:tinf_results_1} but for the other 23 MW satellites in our sample.
    }
    \label{fig:tinf_results_2}
\end{figure*}

In Figures \ref{fig:tinf_results_1} and \ref{fig:tinf_results_2} we show the infall time likelihood for each of the 47 Galactic satellites studied here.
The fiducial result is shown by the solid red curve and includes uncertainties in the measured position and velocity of satellites as well as in the MW potential. To highlight the effect of observational uncertainties, we also show the infall time likelihood inferred using the most likely measure phase-space positions of satellites and the most likely MW mass profile (for more details see Section \ref{sec:obs_data}). The two PDFs are generally in good agreement with each other (e.g. Sagittarius I, LMC and Draco I in figure \ref{fig:tinf_results_1}) and indicate the observational errors are not a significant driver of infall time uncertainties, however for some satellites adding the measurement errors makes a significant difference (e.g. Sextans, Fornax and Grus I in figure \ref{fig:tinf_results_1}, Pisces II and Tucana II in \ref{fig:tinf_results_2}).

\bsp	
\label{lastpage}

\end{document}